\documentclass[
,showpacs ,amssymb,superscriptaddress,aps,dvipdfmx
]{revtex4}
\usepackage{graphicx}
\usepackage{color}
\input{epsf}

\usepackage{amsmath,amssymb}
\usepackage{bm}
\usepackage{times}
\usepackage{ulem}

\newcommand{\dalm}{\kern1pt\vbox{\hrule height 0.9pt\hbox{\vrule width 0.9pt
\hskip 2.5pt\vbox{\vskip 5.5pt}\hskip 3pt\vrule width 0.3pt}\hrule height 0.3pt}
\kern1pt}

\begin{document}



\title{Determination of properties of protoneutron stars toward black hole formation via gravitational wave observations}

\author{Hajime Sotani}
\email{sotani@yukawa.kyoto-u.ac.jp}
\affiliation{Division of Science, National Astronomical Observatory of Japan, 2-21-1 Osawa, Mitaka, Tokyo 181-8588, Japan}

\author{Kohsuke Sumiyoshi}
\affiliation{National Institute of Technology, Numazu College, Ooka 3600, Numazu, Shizuoka 410-8501, Japan}

\date{\today}

\begin{abstract}
We examine the frequencies of the fundamental ($f$-) and the gravity ($g_i$-) mode in gravitational waves from accreting protoneutron stars (PNSs) toward black hole formation. For this purpose, we analyze numerical results of gravitational collapse of massive stars for two different progenitors with three equations of state.  We adopt profiles of central objects obtained from the numerical simulations by solving general relativistic neutrino-radiation hydrodynamics under the spherical symmetry.   Using a series of snapshots as a static configuration at each time step, we solve the eigenvalue problem to determine the specific frequencies of gravitational waves from the evolving PNSs with accretion from core bounce to black hole formation by the relativistic Cowling approximation.  We find that the frequency of the $f$-mode gravitational waves can be expressed as a function of the average density of the PNS at each time step almost independently of the progenitor models in accord with the previous studies on ordinary PNSs.  We also find that the ratio of the $g_1$-mode frequency to the $f$-mode frequency is insensitive to the progenitor, but strongly depends on the compactness of the PNSs.  Having a simultaneous observation of the $f$- and $g_1$-mode gravitational waves, it would provide the information of the average density and compactness of PNS, which can reconstruct the evolution of mass and radius of the PNSs as a constraint on the equation of state at high densities.  Furthermore, we show that the highest average density of the PNS with the maximum mass can be estimated with the detection of the $f$-mode gravitational waves together with the help of the neutrino observations.  
\end{abstract}

%
\maketitle


\section{Introduction}
\label{sec:I}

The gravitational waves are becoming a new tool to hear the information from the Universe. In particular, the signals from the black hole binary mergers and the neutron star binary merger have been already detected by the LIGO (Laser Interferometer
Gravitational-wave Observatory) Scientific Collaboration and Virgo Collaboration.~\cite{GW1,GW2,GW3,GW4,GW5,GW6}. In addition, not only the Japanese gravitational wave detector, KAGRA~\cite{aso13}, will operate soon together with the advanced LIGO and advanced Virgo, but also the discussion for the third-generation detectors, such as the Einstein Telescope and Cosmic Explorer~\cite{punturo,CE}, have already been started. So, in the near future, the number of gravitational wave events would increase more and more. Coming next to the merger of compact object binary, supernova explosions must be another candidate as a gravitational wave source.

Studies of the gravitational waves from core-collapse supernova have been mostly made based on numerical simulations of collapse and bounce of central core in massive stars (e.g.,~\cite{Murphy09,MJM2013,Ott13,CDAF2013,Yakunin15,KKT2016,Andresen16,OC2018}). For example, the recent three-dimensional (3D) numerical simulations for the core-collapse supernova show the existence of the gravitational wave signals associated with the oscillations of PNS (or core region) \cite{KKT2016,RMBVN19,Andresen19}, whose frequency increases with time from a few hundred Hz up to $\sim$ kHz in the postbounce phase. This signal is considered either to arise from the fundamental ($f$-mode) oscillations of the PNS \cite{MRBV2018,SKTK2019}, the gravity ($g$-mode) like oscillations of the whole region inside the shock radius \cite{TCPF2018,TCPOF2019}, or to be the evidence of the Brunt-V\"{a}is\"{a}l\"{a} frequency at the PNS surface (the so-called surface $g$-mode) \cite{MJM2013,CDAF2013}. Since the Brunt-V\"{a}is\"{a}l\"{a} frequency is a local value, which is not associated with the global oscillations of the PNS, the gravitational wave signals in numerical simulations may be a result of the $f$-mode oscillations of the PNS or the $g$-mode like oscillations inside the shock radius. In addition to the PNS oscillations, the 3D simulations show another gravitational wave signal, whose frequencies are close to the modulation frequency of the standing accretion-shock instability, $\sim 100$ Hz \cite{KKT2016,Andresen16,OC2018,RMBVN19}.

In order to extract a physical property of compact objects from observations, asteroseismology is a powerful technique~\cite{KS1999}, which is a similar way in seismology for the Earth and helioseismology for the Sun. Since the specific oscillation frequencies of objects strongly depend on the interior information, one may inversely derive the interior information by observing the oscillation frequencies. In fact, one could constrain the crustal properties by identifying the quasi-periodic oscillations observed in the magnetar giant flares as the crustal torsional oscillations in neutron stars (e.g.,~\cite{SW2009,GNHL2011,SNIO2012,SIO2016}). In the similar way, the direct detection of the gravitational waves from the compact objects enables us to obtain the information about the radius, mass, and equation of state (EOS) of the source objects (e.g.,~\cite{AK1996,AK1998,STM2001,SH2003,SYMT2011,PA2012,DGKK2013}).

Compared to the extensive studies of asteroseismology for cold neutron stars, the number of the studies of the PNS asteroseismology is relatively limited~\cite{FMP2003,Burgio2011,FKAO2015,ST2016,Camelio17,SKTK2017,MRBV2018,TCPF2018,TCPOF2019,SKTK2019}. This may partially come from the difficulty for preparing the PNS model as a background for the linear analysis. That is, the cold neutron star models can be constructed by adopting the relation between the density and pressure, while one has to prepare the radial distributions of the density, pressure, electron fraction, and entropy per baryon to construct the PNS models. However, such radial distributions can be determined, in principle, only after the numerical simulations of the core-collapse supernova starting from the gravitational collapse of massive stars. Hence, the study of PNS asteroseismology requires construction of PNS models extracted from numerical simulations in an appropriate way.  Owing to the long-standing difficulties of the numerical simulations of supernova explosions, the number of studies of the PNS asteroseismology increases only gradually.

We focus on the asteroseismology of accreting PNS toward black hole formation from massive stars in the current study.  In a certain class of massive stars with a large compactness \cite{OConnor11}, a nascent PNS becomes increasingly massive due to the continuation of mass accretion after core bounce \cite{Sumiyoshi06}.  The black hole formation occurs through dynamical collapse when the mass of PNS eventually reaches the critical mass.  Extensive studies of such black-hole-forming collapse of massive stars show that the evolution of central object toward the black hole formation depends on the properties of progenitor and the EOS~\cite{Sumiyoshi07,Sumiyoshi08}.  The former determines the accretion rate through density profiles and the latter determines the maximum mass of accreting PNS.  It is to be noted that the black-hole-forming collapse is associated with the short burst of energetic neutrinos, which ends at the black-hole formation.
The neutrino emission with increasing average energies from the accreting PNS continues for $\sim$1 sec, depending on the progenitor and the EOS, and is terminated at the black hole formation.  Hence, future detection of such neutrino bursts will provide us with the information of the accreting PNS toward black hole formation \cite{Nakazato10} and will be helpful for the identification of gravitational waves.  We remark that the correlation between the gravitational waves and neutrino signals have been discussed for the  core-collapse supernova (e.g.,~\cite{Yokozawa15,Kuroda17,Takiwaki18,WS2019}).  In addition, a study of dynamics and multi-messenger signals from black hole forming case has been made in multi-dimensional simulations recently \cite{Pan18,Kuroda18}.

In this study, we examine the eigenfrequencies of gravitational waves in profiles of PNS taken from the numerical simulations of black hole formation from massive stars under the spherical symmetry \cite{Sumiyoshi07,Sumiyoshi08,Nakazato10}.  We adopt PNS models for two progenitors with three EOSs to show the universal characters of gravitational waves.  Each model represents the sequence of accreting PNS with increasing mass and covers the time evolution from the core bounce to the black hole formation.  Since the density becomes high as the mass of PNS increases, we can explore the change of eigenfrequencies according to the rise of average density.  Note that these models are provided together with the detailed evaluation of neutrino emission during this evolution.  We utilize the neutrino information to analyze the eigenfrequencies of the accreting PNS in future gravitational wave detection expecting that the neutrino detectors such as Super-Kamiokande will record the neutrino emission until the termination of neutrino emission.  This may enable us to extract the highest density of the PNS reached at the moment of black hole formation identified by the timing of the termination of neutrino emission.   Since the maximum mass of PNS is determined by the EOS, the simultaneous observations of the gravitational waves and neutrinos may provide a probe of EOS at high densities.

For the PNS asteroseismology in supernovae, two different approaches have been adopted so far. One approach is the way to construct the PNS model, whose surface is determined by a specific surface density~\cite{ST2016,SKTK2017,MRBV2018,SKTK2019}, while another approach is that the whole region inside the shock radius is considered as a background~\cite{TCPF2018,TCPOF2019}. For the first approach, one has to select the surface density and the eigenfrequencies of the PNS in the early phase (up to $\sim 500$ ms) after core bounce depend on the selection of this density~\cite{MRBV2018,SKTK2019}, but the boundary condition imposed at the surface is simple, i.e., the Lagrangian perturbation should be zero. Since this boundary condition is the same as in the standard asteroseismology, one can easily classify the eigenfrequencies by counting the radial nodal numbers in the eigenfunctions. On the other hand, for the second approach, one can avoid the uncertainty in the selection of surface density, which is an advantage for this approach.  The background models are assumed to be spherically symmetric by averaging the angular direction based on multi-dimensional simulations even though the matter motion in the outer region with a low density may be relatively violent.  In addition, one has to impose the boundary condition that the radial displacement should be zero at the shock radius~\cite{TCPF2018,TCPOF2019} to determine the eigenfrequency. This is a completely different mathematical problem from the standard asteroseismology, where the mode classification should be done with another definition.  

In this study we take the first approach by adopting the profiles of PNSs from numerical simulations. We remark that we do not need to perform an angular average for preparing the PNS model as in the previous studies for the PNS asteroseismology since we adopt results of numerical simulations under the spherical symmetry in this study.  We identify the fundamental modes such as $f$-mode of the gravitational wave signal by choosing the surface density to be $10^{11}$ g/cm$^3$ \cite{MRBV2018,SKTK2019}.  The choice of the surface density with $10^{11}$ g/cm$^3$ is supported by the analysis of gravitational waves in the 3D supernova simulation by Kuroda although the eigenfrequencies may depend on the choice \cite{SKTK2019} and there is ambiguity in its determination due to hydrodynamics and numerical resolution.

This paper is arranged as follows.  In section \ref{sec:PNSmodel}, we describe the setup of PNS models and their properties toward the black hole formation.  In section \ref{sec:GW}, we report the eigenfrequencies of gravitational waves and the analysis of fundamental modes during the evolution of accreting PNSs.  We show that the $f$-mode and the ratio of $g_1$-mode to $f$-mode of gravitational waves depends on the average density and the compactness of PNS, respectively, so that one can track the evolution of mass and radius of PNS as a probe of EOS.  We summarize the paper in section \ref{sec:Conclusion}.  
Unless otherwise mentioned, we adopt geometric units in the following, $c=G=1$, where $c$ denotes the speed of light, and the metric signature is $(-,+,+,+)$.

\section{PNS Models}
\label{sec:PNSmodel}

We prepare PNS models as a background for the linear analysis by adopting the profiles from the numerical studies of core-collapse and bounce of massive stars under the spherical symmetry.  The numerical simulations have been performed by solving the general relativistic neutrino-radiation hydrodynamics which handles hydrodynamics and neutrino transfer by Boltzmann equation in general relativity \cite{Yamada97,Yamada99,Sumiyoshi05}.  The Boltzmann equation is directly solved by treating the multi-angle and multi-energy neutrino distributions.  The collision term handles the basic neutrino reaction rates \cite{Bruenn85} for emission, absorption and scattering with nucleons and nuclei, scattering with electrons and positrons, pair creations and annihilations including the nucleon-nucleon bremsstrahlung and plasmon process \cite{Sumiyoshi05}.  The numerical code has been applied to the study of core-collapse supernovae \cite{Sumiyoshi05} and the supernova neutrino database \cite{Nakazato13} by following the time evolution from the gravitational collapse of iron core of massive stars, core bounce and shock propagation.  It has been also applied to the black hole formation from massive stars by following the evolution of nascent PNS with intense accretion after core bounce.  The dynamics of accreting PNS toward the black hole formation with neutrino emission has been reported in Ref.~\cite{Sumiyoshi06,Sumiyoshi07,Sumiyoshi08}.  The numerical code uses the metric given by 
\begin{equation}
  ds^2 = -e^{2\Phi(t,m_b)} dt^2 + e^{2\Lambda(t,m_b)} dm_b^2 + r^2(t,m_b)(d \theta^2 + \sin^2\theta\, d \phi^2), \label{eq:metric0}
\end{equation}
where $t$ and $m_b$ are the coordinate time and the baryon mass coordinate, respectively \cite{MS64}. We note that $m_b$ is associated with the circumference radius $r$ via the baryon mass conservation.  The metric components, $\Phi(t,m_b)$ and $\Lambda(t,m_b)$, are evolved in the numerical simulations and are provided with hydrodynamical variables for PNS configurations.  (See \cite{Yamada97} for definitions in detail.)  The numbers of grids for radial mass coordinate, neutrino angle and energy are 255, 6 and 14, respectively.  The radial grids of the mass coordinate are arranged in a non-uniform manner to cover the central object and matter accretion.

In order to see the dependence on the progenitor models in the gravitational wave signals from the PNSs after core-bounce, we focus on two different progenitor models, i.e., a $40M_\odot$ star based on Ref.~\cite{WW95}  and a $50M_\odot$ star based on Ref.~\cite{TUN07}, which are hereafter refereed to as W40 and T50, respectively. 
These progenitors have large values for the compactness \cite{OConnor11}, which expresses the extent of dense profiles, and lead to the stall of shock wave with intense accretion onto the PNS just born after the core bounce without any possibility of explosion.  
This situation is in contrast to the case of progenitors with small values for the compactness, with which the explosion dynamics is explored and ordinary PNSs are born without significant accretion.  (See \cite{OConnor11,Sukhbold16,Horiuchi18}, for example, on the boundary between the birth of PNS and black hole.)

For each progenitor model, we adopt three different EOSs, i.e., Shen EOS \cite{Shen_EOS}, LS180, and LS220 \cite{LS_EOS}. These sets of supernova EOS provide thermodynamical quantities of hot and dense matter as well as mass fractions in mixture of neutrons, protons, $^4$He and a representative species of nuclei with electrons, positrons and photons.  Shen EOS is based on the relativistic mean field theory with the TM1 nuclear interaction, where the incompressibility $K_0=281$ MeV, the symmetry energy $S=36.9$ MeV and the slope parameter of the nuclear symmetry energy $L=110.8$ MeV, while LS180 (LS220) is constructed with the compressible liquid drop model using the functional form of energy with $K_0=180$ MeV (220 MeV), $S=29.3$ MeV and $L=73.8$ MeV. These sets cover different stiffness of EOSs having the maximum mass of cold neutron stars with $2.2M_\odot$ for Shen and $1.8M_\odot$ ($2.0M_\odot$) for LS180 (LS220).  (See Ref. \cite{Sumiyoshi04,Oertel17} for the detailed comparison of EOS.)  Note that PNSs are proton rich and hot, being different from cold neutron stars.  Hence, they have different maximum masses from the cold neutron stars.  
We remark that we consider not only Shen and LS220 but also LS180, even though LS180 is excluded by the $2M_\odot$ observations~\cite{Demorest2010,Antoniadis2013}, because we would like to examine the dependence of the gravitational wave spectra on the stiffness of EOS.
The evolution of PNS toward the black hole formation and properties of neutrino emissions for these progenitors and EOS can be found in Ref.~\cite{Sumiyoshi07,Sumiyoshi08}.  We denote Shen, LS180 and LS220 for three EOS sets in combination with the progenitor name, W40 and T50, although we will not consider the combination of LS220 with T50. This is because LS180 and LS220 are similar to each other while LS180 and Shen represent an extreme case of soft and stiff EOS, respectively.

\begin{figure}[tbp]
\begin{center}
\begin{tabular}{ccc}
\includegraphics[scale=0.43]{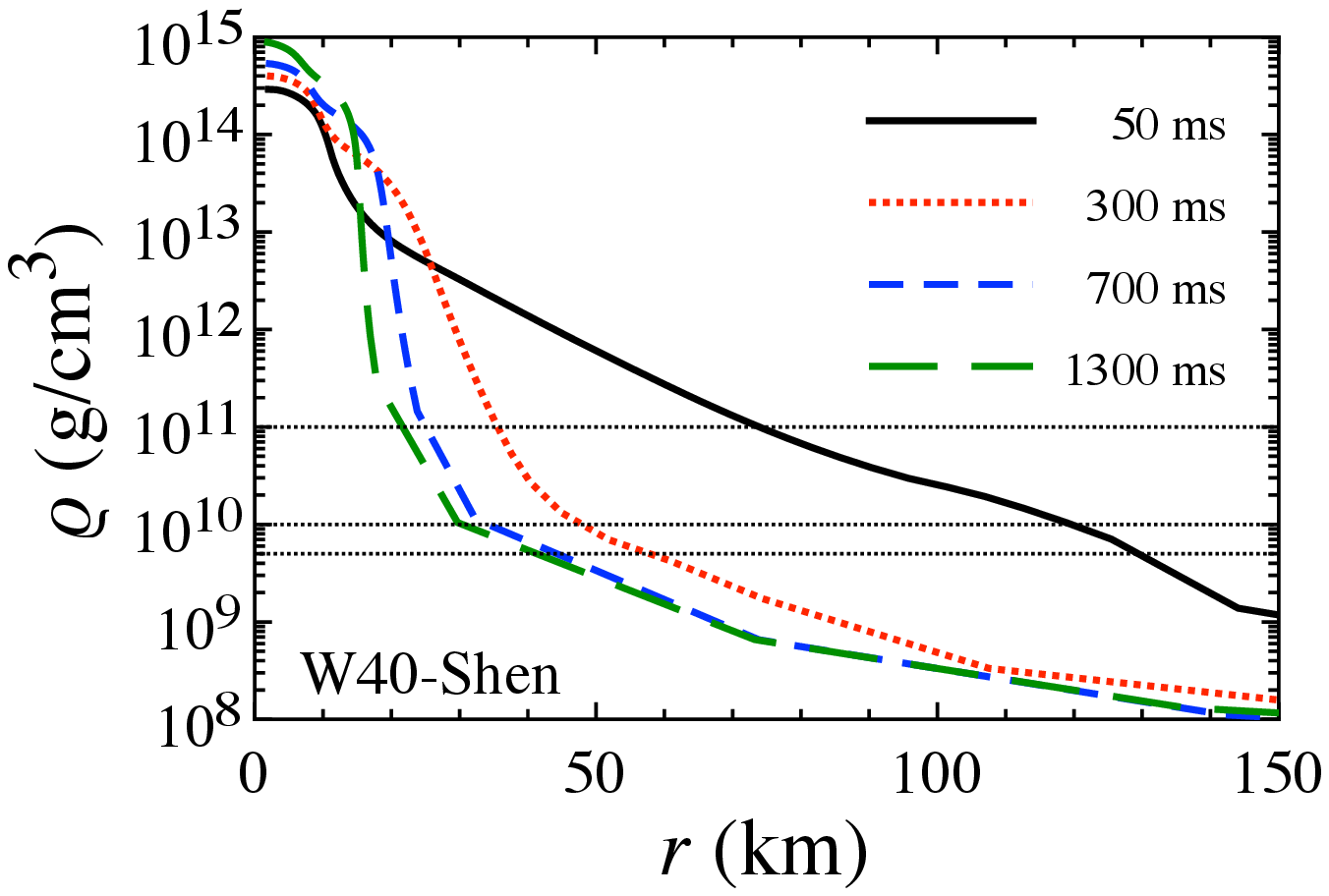} &
\includegraphics[scale=0.43]{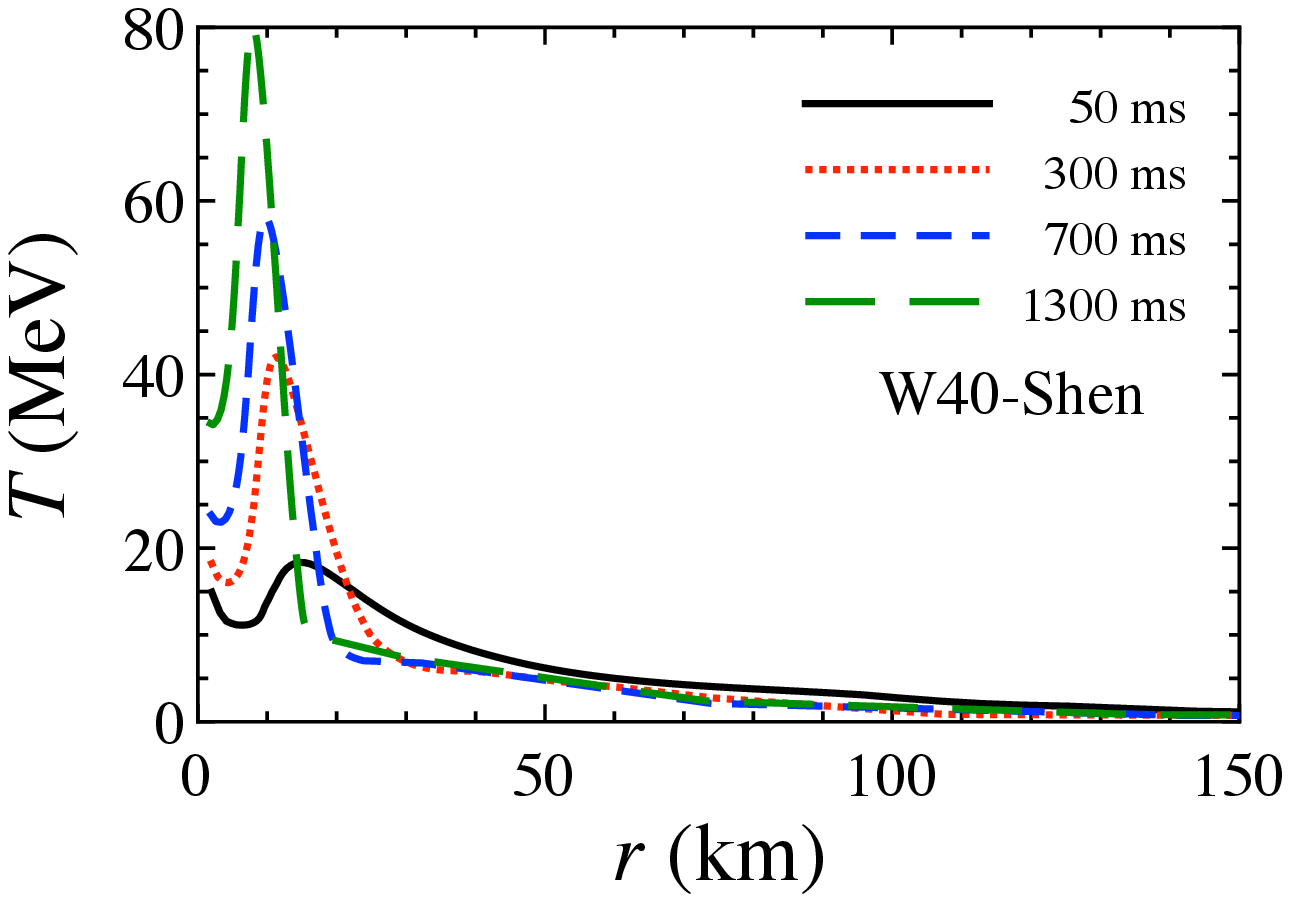} &
\includegraphics[scale=0.43]{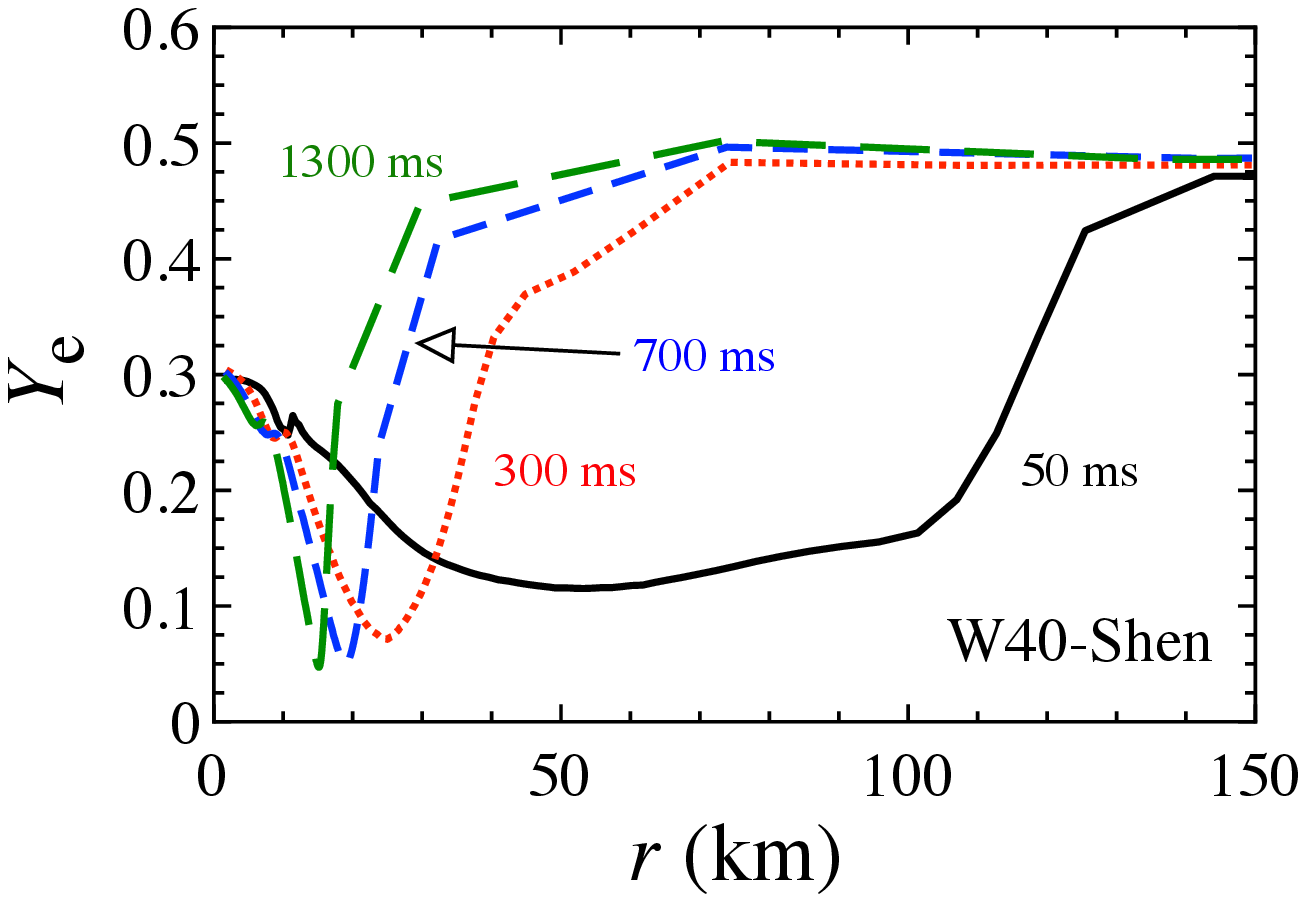} 
\end{tabular}
\end{center}
\caption{
Rest mass density (left panel), temperature (middle panel), and electron fraction profiles (right panel) are shown for the PNS model of W40-Shen at $T_{\rm pb}=50$, 300, 700, and 1300 ms. For reference, $\rho_s=10^{11}$, $10^{10}$, and $5\times 10^9$ g/cm$^3$ are shown by the horizontal lines in the left panel. 
}
\label{fig:rho-W40Shen}
\end{figure}

\begin{figure*}[tbp]
\begin{center}
\begin{tabular}{cc}
\includegraphics[scale=0.5]{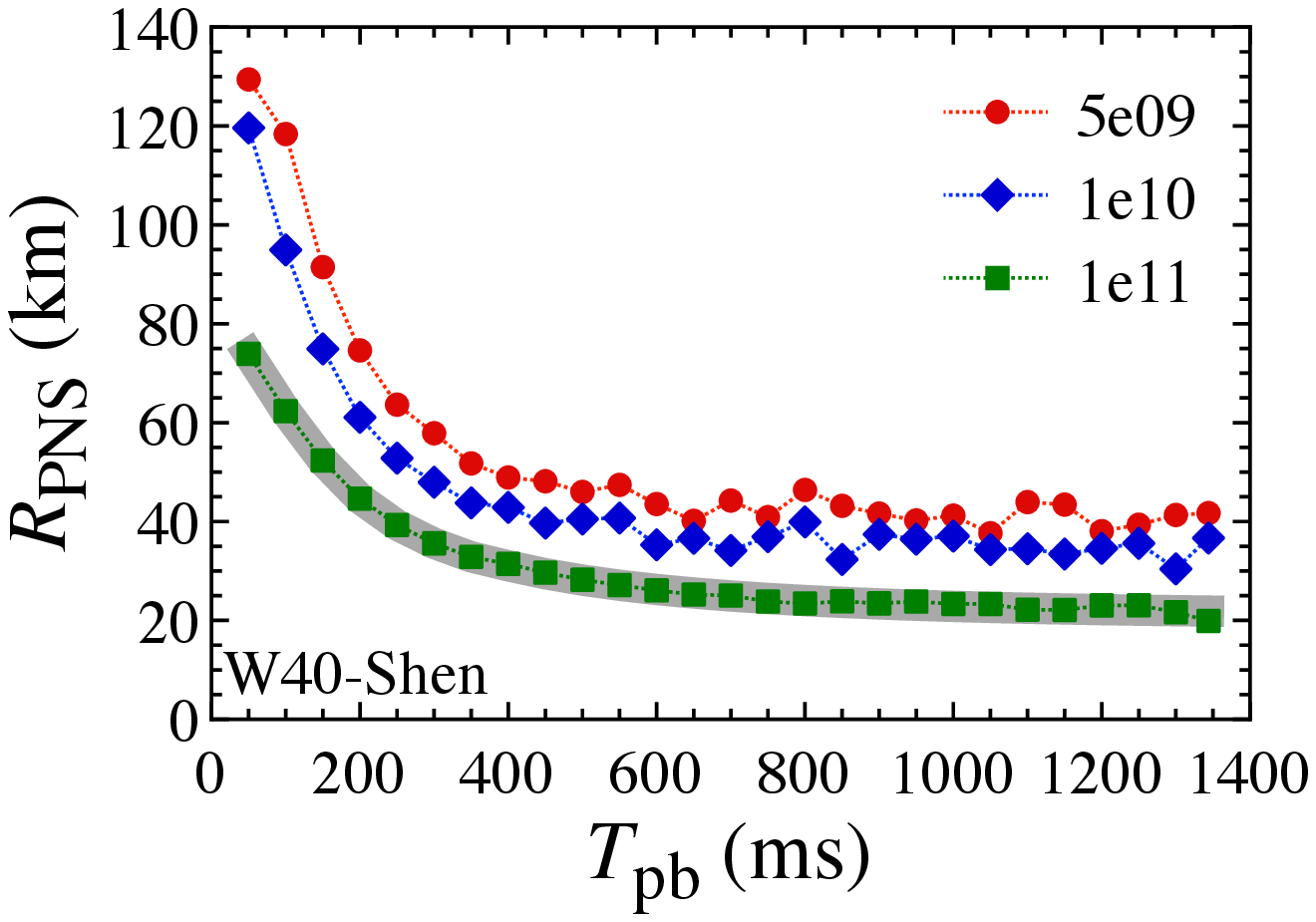} &
\includegraphics[scale=0.5]{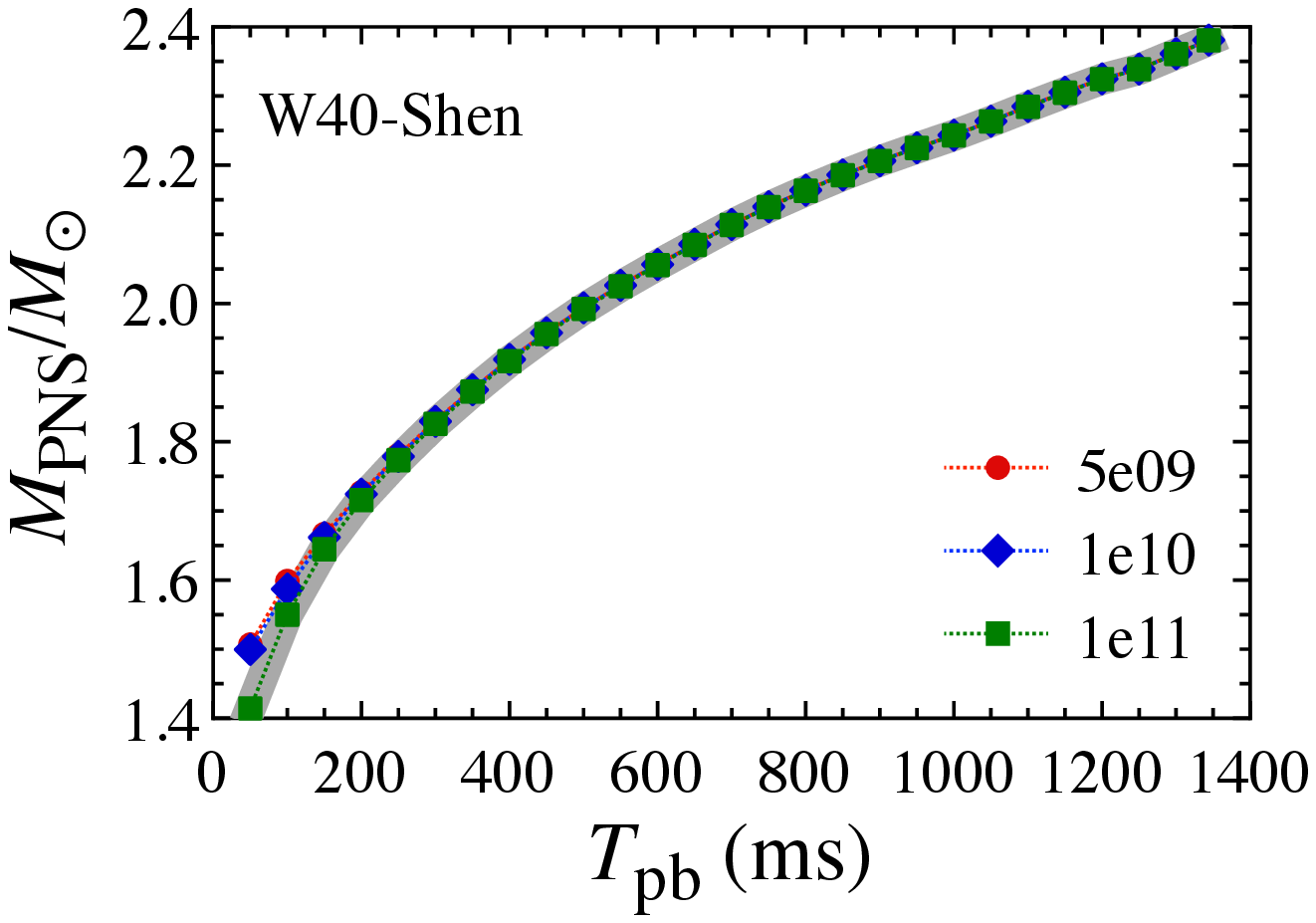}  
\end{tabular}
\end{center}
\caption{
Evolution of the radius (left) and the gravitational mass (right) of PNS models of W40-Shen with different surface density, where the circles, diamonds, and squares correspond to $\rho_s=5\times 10^9$, $10^{10}$, and $10^{11}$ g/cm$^3$. In addition, for the case with $\rho_s=10^{11}$ g/cm$^3$, the radius by fitting with Eq. (\ref{eq:fitR}) for $T_{\rm pb}\ge 400$ ms is shown by the solid-thick line in the left panel, where the radius for $T_{\rm pb}<400$ ms is the same as the data with squares without fitting. In the right panel, the gravitational mass inside the radius shown by the solid-thick line in the left panel is also shown by the solid-thick line. 
}
\label{fig:Mt-W40Shen}
\end{figure*}

We prepare the PNS profiles from the five sets of numerical simulations (W40 with Shen, LS180 and LS220 and T50 with Shen and LS180.  See also Table \ref{tab:fitR}).  We adopt the time sequence of profiles at every 50 ms after core bounce up to the moment of re-collapse to black hole for each model and set the radius of PNS at a specific surface density, $\rho_s$.  In Fig.~\ref{fig:rho-W40Shen}, we show the radial profile of rest-mass density (left panel), temperature (middle panel), and electron fraction (right panel) for the PNS models of W40-Shen at $T_{\rm pb} = 50$, 300, 700, and 1300 ms.  Hereafter, we adopt $T_{\rm pb}$ as a time after the core bounce in the unit of millisecond, i.e., $T_{\rm pb}=0$ at the core bounce. We also show the lines with $\rho_s=10^{11}$, $10^{10}$, and $5\times 10^9$ g/cm$^3$ to examine the dependence of the radius on the choice of $\rho_s$ in the left panel.  One may notice coarse profiles near $\rho_s$ for 700, and 1300 ms due to a low resolution for extended surface.  This is because the numerical simulations have been done using the baryon mass coordinate, which is lagrangian.  The surface of PNS can be under-resolved due to the limited grid when the accretion rate becomes small at late stages while the outer layers have more grids.  In order to check the dependence on $\rho_s$ and the resolution, we show in Fig.~\ref{fig:Mt-W40Shen} the evolutions of the PNS radius and mass defined with $\rho_s=10^{11}$, $10^{10}$, and $5\times 10^{9}$ g/m$^3$ for the model of W40-Shen.  The PNS radius decreases and the PNS mass increases due to the mass accretion.  One can see that the PNS radii are smooth up to $\sim 500$ ms, but the radii with $\rho_s=5\times10^{9}$ and $10^{10}$ g/cm$^3$ show wiggling behavior at later stages.  The PNS radius with $\rho_s=10^{11}$ g/cm$^3$ behaves rather smoothly.  This also indicates the ambiguity of radii at the surface with low densities.  In contrast, the time evolution of PNS mass is smooth and insensitive to the choices of $\rho_s$ as shown in the right panel of Fig.~\ref{fig:Mt-W40Shen}.  In order to provide the smooth values of radii for the evaluation of eigenfrequencies, we fit the PNS radius with $\rho_s=10^{11}$ g/cm$^3$ for $T_{\rm pb}\ge 200$ ms by adopting the fitting formula proposed in Ref. \cite{SKJM06}, i.e.,
\begin{equation}
  R_{\rm PNS}(T_{\rm pb}) = R_i\left[1+\left\{1-\exp\left(-\frac{T_{\rm pb}}{\tau}\right)\right\}\left(\frac{R_i}{R_f}-1\right)\right]^{-1}, 
   \label{eq:fitR}
\end{equation}
where $R_i$ and $R_f$ denote the radii of the PNSs at $T_{\rm pb}=0$ and $\infty$, respectively, while $\tau$ is a typical time scale for the evolution of the PNSs in milliseconds. These parameters in the fitting formula for the PNS models considered in this study are shown in Table \ref{tab:fitR}. For the case of W40-Shen, the fitting line given by Eq. (\ref{eq:fitR}) is also shown in Fig.~\ref{fig:Mt-W40Shen} with the thick-solid line for $T_{\rm pb}\ge 400$ ms, where the line for $T_{\rm pb}< 400$ ms shows the raw data without fitting.

\begin{table}
\centering
\caption{
Parameters in the fitting formula given by Eq. (\ref{eq:fitR}) for the PNS models considered in this study, using the data for $T_{\rm pb}\ge 200$ ms. 
}
\begin{tabular}{cccccc}
\hline\hline
  EOS & Model & $R_i$ (km) & $R_f$ (km) & $\tau$ (ms) & $T_{\rm BH}$ (ms) \\
\hline
  Shen & W40 &  158.33 &  21.25  & 396.08 & 1344 \\ 
            & T50 & $8.086\times 10^6$ & 22.98 & 288.72 & 1509 \\ 
 LS180 & W40 &  9308.31 & 13.52 & 492.89 & 558 \\ 
           &  T50 &   $6.985\times 10^7$ &  10.04 & 707.69 & 505 \\ 
 LS220 & W40 & $1.654\times 10^8$ & 17.51 & 327.96 & 783 \\ 
\hline\hline
\end{tabular}
\label{tab:fitR}
\end{table}

\begin{figure*}[tbp]
\begin{center}
\begin{tabular}{cc}
\includegraphics[scale=0.42]{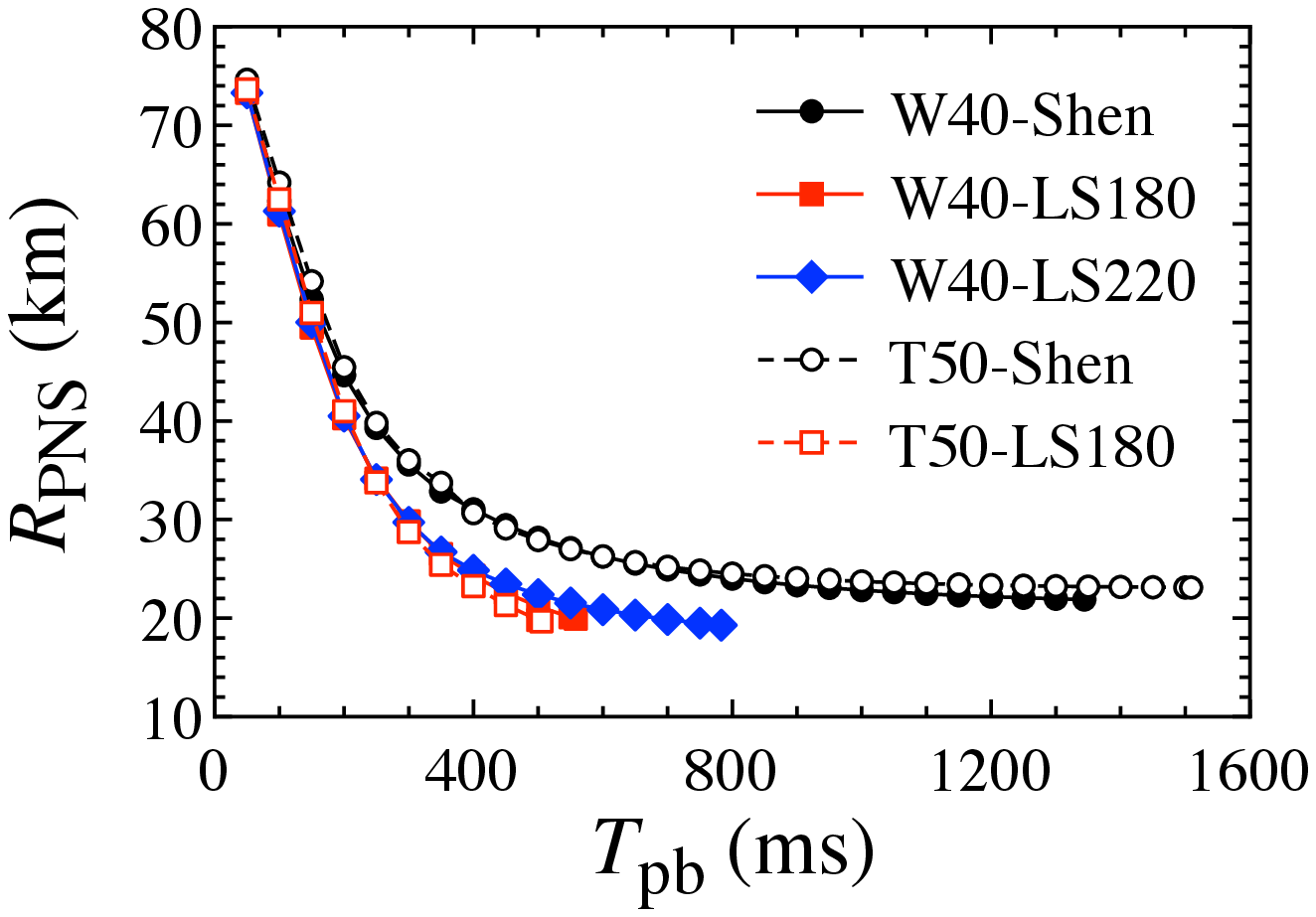} &
\includegraphics[scale=0.42]{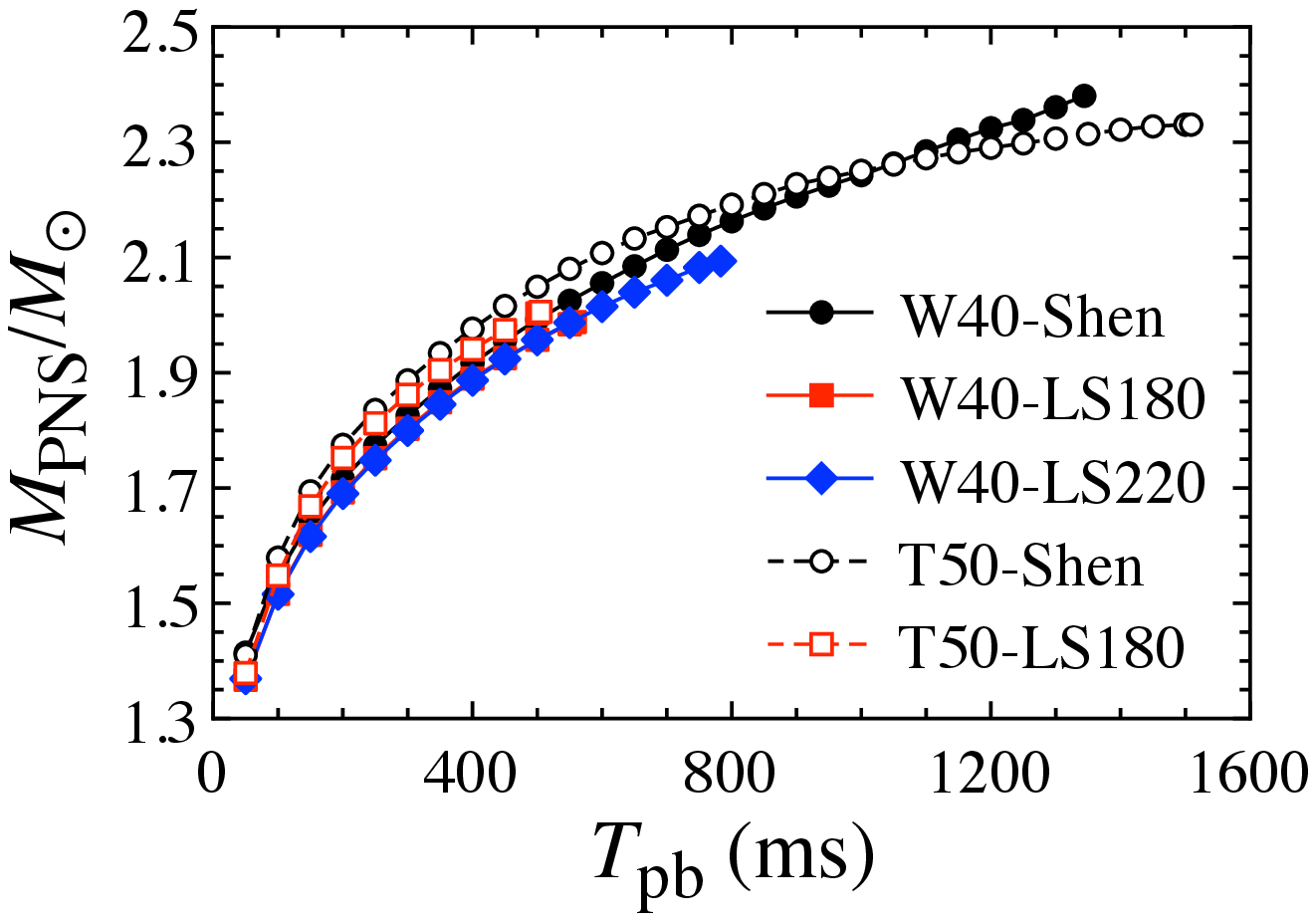}  
\includegraphics[scale=0.42]{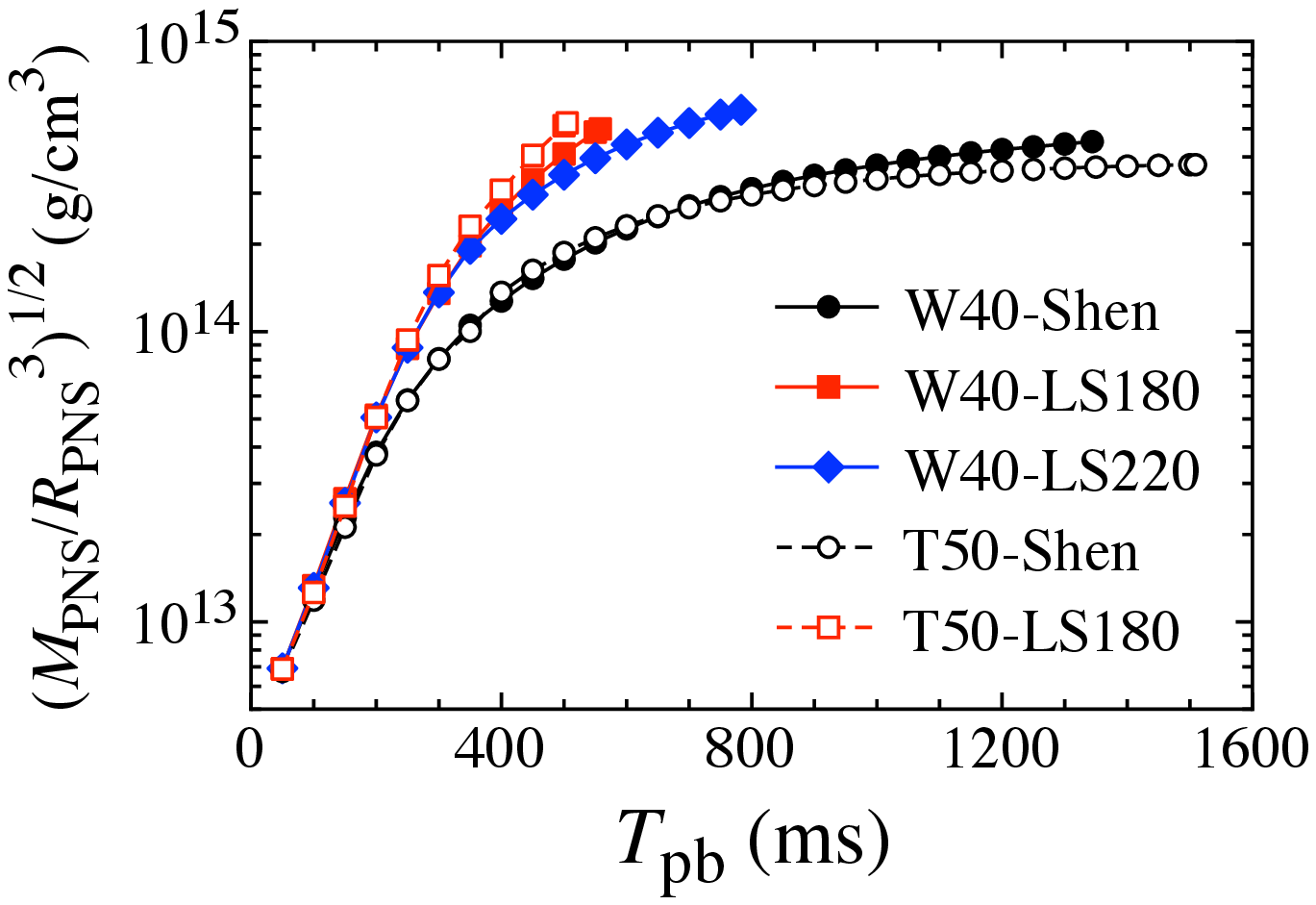}  
\end{tabular}
\end{center}
\caption{
Evolution of the PNS radius (left), the gravitational mass (center), and the average density (right) for various PNS models, where the surface density is fixed to be $10^{11}$ g/cm$^3$.
}
\label{fig:PNSt}
\end{figure*}

In the following analysis, we consider only the case of PNS models with $\rho_s=10^{11}$ g/cm$^3$.  This choice is supported by the finding that the pattern of gravitational wave signals in the general relativistic 3D simulations is in a good agreement in the $f$-mode oscillations of the PNS model with $\rho_s=10^{11}$ g/cm$^3$ \cite{SKTK2019}.  It is, however, necessary to remind that the eigenfrequencies may depend on the choice of $\rho_s$ \cite{SKTK2019}.  The fitting formula, Eq.~(\ref{eq:fitR}), with the coefficients in Table~\ref{tab:fitR} is adopted for $T_{\rm pb}\ge 400$ ms.  We show in Fig.~\ref{fig:PNSt} the evolution of radius, mass, and the  average density, $M_{\rm PNS}/R_{\rm PNS}^3$ for the five PNS models. In each panel, the rightmost endpoint of each line corresponds to the time for the black hole formation, $T_{\rm BH}$, which must be determined observationally via the neutrino signals \cite{Nakazato10}. The value of $T_{\rm BH}$ for each PNS model is also shown in Table~\ref{tab:fitR}. We remark that the PNS model at $T_{\rm pb}=T_{\rm BH}$ corresponds to that with the maximum mass. Finally, by assuming that the PNS model is in a static equilibrium at each time step, we can prepare the PNS model as a background for considering the linear perturbation analysis. In this case, the metric is written with the spherical coordinate as
\begin{equation}
  ds^2 = -e^{2\Phi(r)} dt^2 + e^{2\Lambda(r)} dr^2 + r^2(d \theta^2 + \sin^2\theta\, d \phi^2). \label{eq:metric1}
\end{equation}

\section{Protoneutron star asteroseismology}
\label{sec:GW}

We perform the linear perturbation analysis of the PNS models described above. Once the PNS models are given, how to determine the eigenfrequencies of PNS is the same as in Ref. \cite{SKTK2019}. That is,  for simplicity, we assume the relativistic Cowling approximation, i.e., the metric perturbations are neglected during the fluid oscillations. In this case, the perturbation equations can be derived by linearizing the energy momentum conservation law. The concrete equation system is given by Eqs. (23) and (24) in Ref.~\cite{SKTK2019}. By imposing appropriate boundary conditions, it eventually becomes an eigenvalue problem with respect to the eigenvalue of $\omega$. Then, the eigenfrequency of gravitational wave is determined by $\omega/(2\pi)$. As the boundary conditions, one has to impose the regularity condition at the stellar center, while the Lagrangian perturbation of pressure should be zero at the PNS surface. 
We remark that the linear analysis with the Cowling approximation does not provide information on  damping times of gravitational waves. This information may be important to determine which oscillating mode is competitive in radiating energy and therefore can possibly be observed as discussed in Refs.~\cite{FMP2003,Burgio2011} (see also discussion in Section \ref{sec:Conclusion}).
We also remark that the Cowling approximation may approximate well for the case of PNSs at least in early phase after core bounce, because the relativistic effect in such a phase is not so strong due to the less compactness of PNSs. In fact, according to the results in Ref.~\cite{TCPOF2019}, where the metric perturbations are still partially taken into account, the frequency with the Cowling approximation deviates from that partially including metric perturbations with time. Probably, the deviation between the frequency with and without the Cowling approximation may become at most $\sim 20\%$, which is the results for the cold neutron stars, but may correspond to the final state for the evolution of PNSs~\cite{Yoshida97}.

\begin{figure*}[tbp]
\begin{center}
\begin{tabular}{cc}
\includegraphics[scale=0.5]{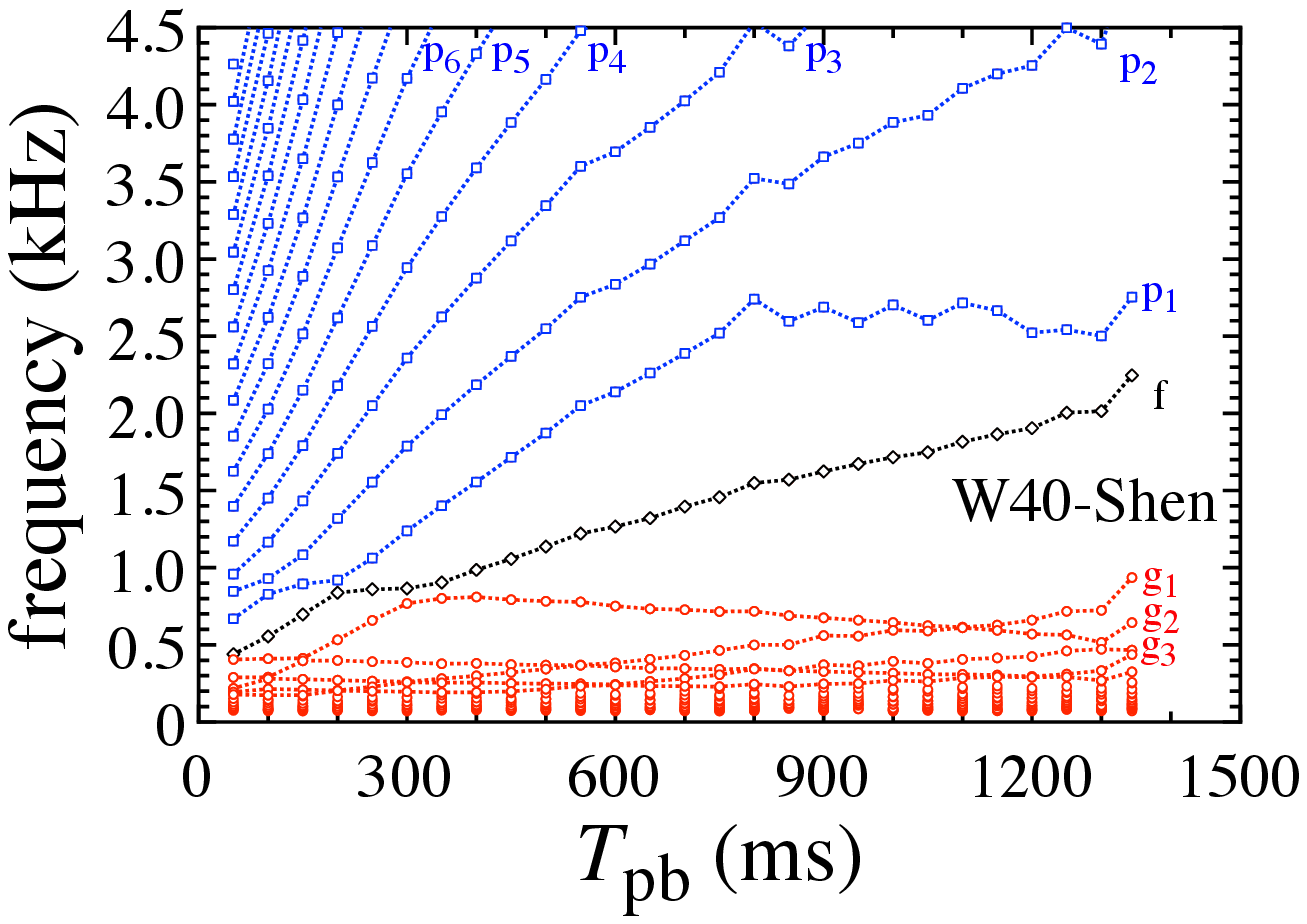} &
\includegraphics[scale=0.5]{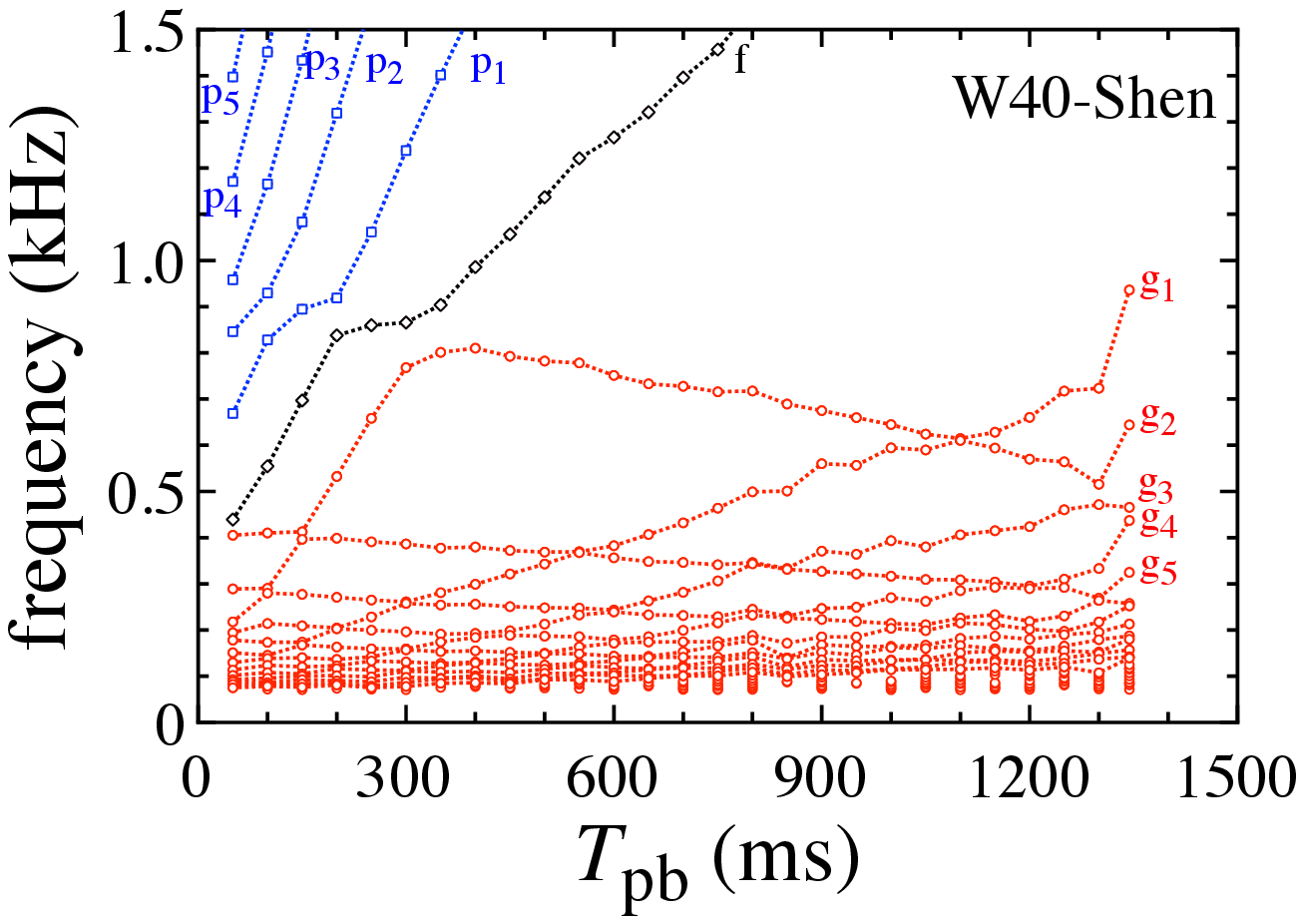}  
\end{tabular}
\end{center}
\caption{
Evolution of the eigenfrequencies for the PNS model of W40-Shen. The right panel is just an enlarged view of the left panel. The $f$-, $p_i$-, and $g_i$-modes are shown with the diamonds, squares, and circles. 
}
\label{fig:f-W40Shen}
\end{figure*}

First, we show the evolution of eigenfrequencies for the PNS model of W40-Shen in Fig.~\ref{fig:f-W40Shen}, where the right panel is just an enlarged view of the left panel. The diamonds, squares, and circles correspond to the frequencies of the $f$-, $p_i$-, and $g_i$-mode gravitational waves. One can clearly observe the phenomena of the avoided crossing. That is, for example, focusing on the $f$-mode frequency, the avoided crossing occurs with $p_1$-mode at $T_{\rm pb}\sim 200$ ms and with $g_1$-mode at $T_{\rm pb}\sim (300-350)$ ms. Due to  the avoided crossing, one can observe the plateau for a while in the evolution of the $f$-mode frequency. The evolutions of eigenfrequencies for the other PNS models are shown in Appendix~\ref{sec:appendix_1}.

\begin{figure}[tbp]
\begin{center}
\includegraphics[scale=0.5]{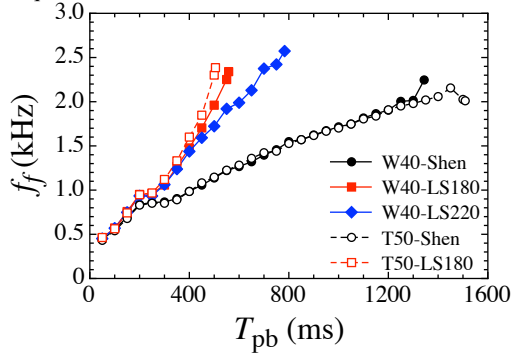} 
\end{center}
\caption{
Evolution of the frequency of the $f$-mode gravitational waves for various PNS models. 
}
\label{fig:fft}
\end{figure}

\begin{figure*}[tbp]
\begin{center}
\begin{tabular}{cc}
\includegraphics[scale=0.5]{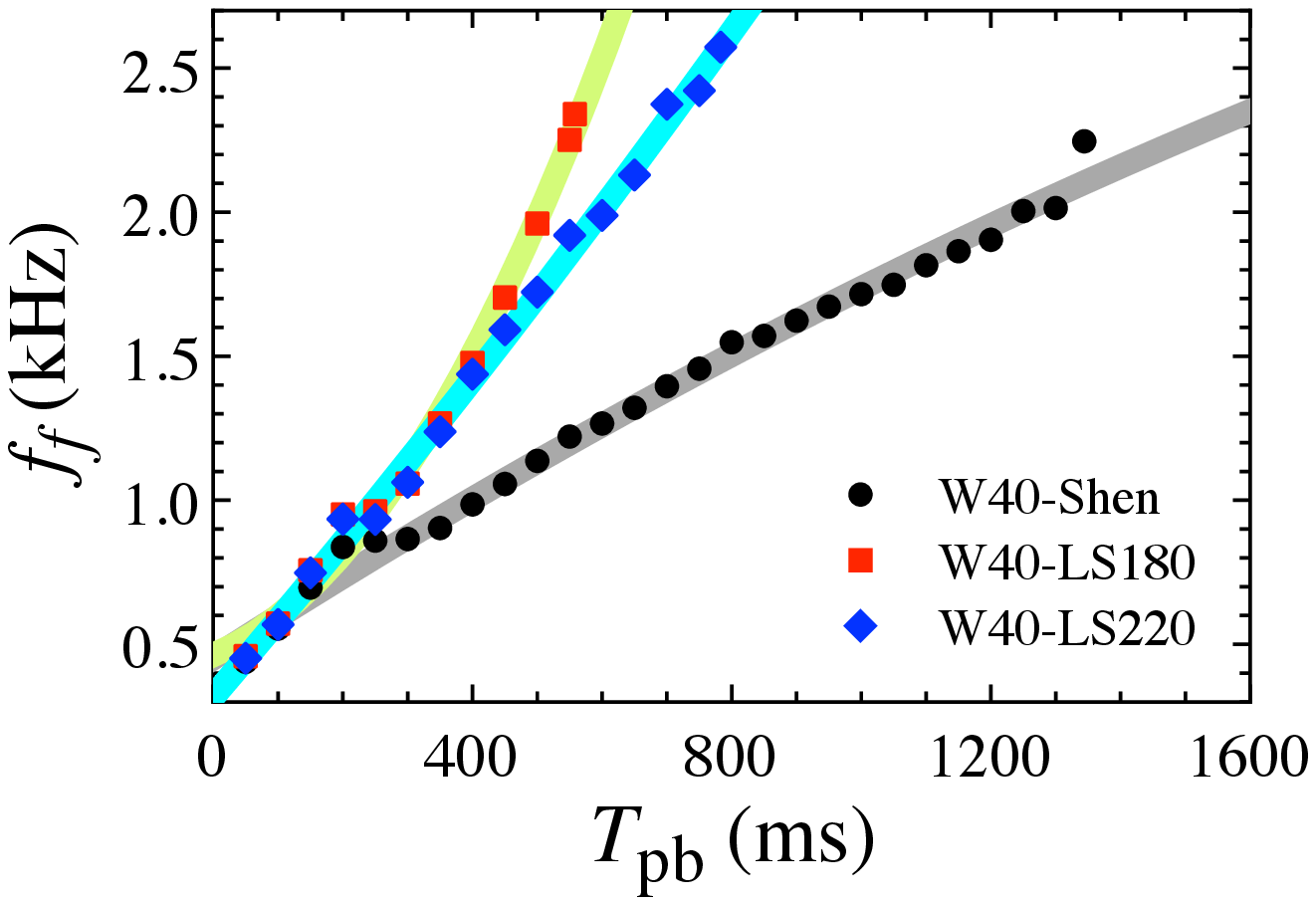} &
\includegraphics[scale=0.5]{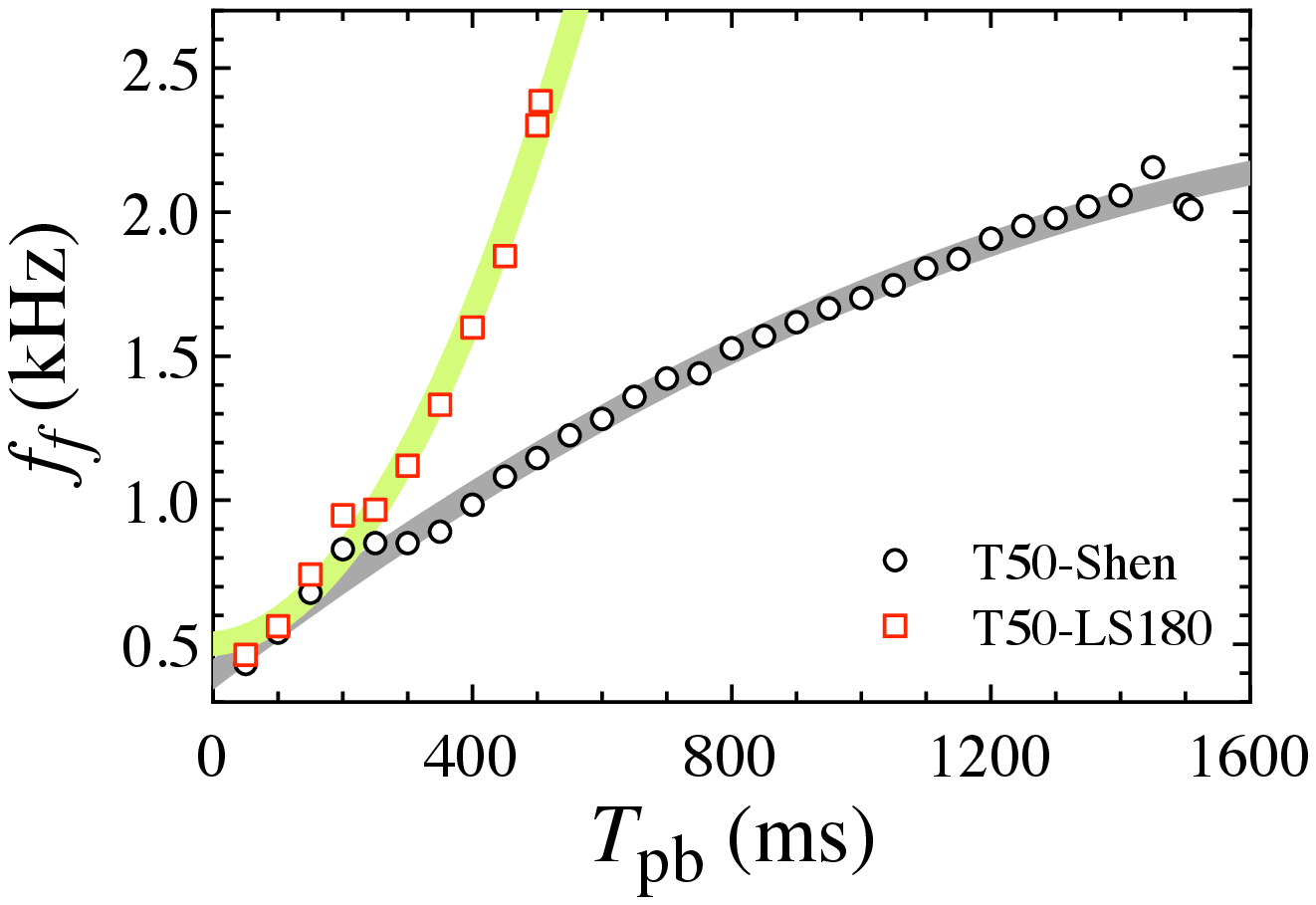}  
\end{tabular}
\end{center}
\caption{
Comparison between $f_f$ obtained by eigenvalue problems (marks) and the expectation with the fitting formula given by Eq. (\ref{eq:fit_fft}) (solid-thick lines). The left and right panels correspond to the cases for the PNS models with W40 and T50, respectively. 
}
\label{fig:fft-fit}
\end{figure*}

Next, we focus on the behavior of the $f$-mode frequency. In Fig.~\ref{fig:fft}, the evolution of $f$-mode frequency is shown for various PNS models. It is interesting to commonly observe the plateau for a while around $T_{\rm pb}\sim (200-300)$ ms.  As a general trend except for the small plateau, we find that the $f$-mode frequency can be fitted well as
\begin{equation}
  f_f ({\rm kHz}) = c_0 + c_1\left(\frac{T_{\rm pb}}{1000\,{\rm ms}}\right)
                          + c_2\left(\frac{T_{\rm pb}}{1000\,{\rm ms}}\right)^2, \label{eq:fit_fft}
\end{equation}
where $c_0$, $c_1$, and $c_2$ are fitting coefficients depending on the PNS models, as shown in Table \ref{tab:fit_fft}. In Fig.~\ref{fig:fft-fit}, the fitted line for each PNS model is shown with the thick-solid line. By using this fitting formula together with the information about $T_{\rm BH}$ determined from the neutrino observations, one could estimate the $f$-mode frequency from the PNS model with the maximum mass (at $T_{\rm pb}=T_{\rm BH}$), even though the $f$-mode gravitational wave at $T_{\rm pb}\sim T_{\rm BH}$ would not be observed directly due to the fact that the frequency of $f$-mode in the final phase is relatively high for observation.  

\begin{table}
\centering
\caption{
Coefficients in the fitting formula given by Eq. (\ref{eq:fit_fft}) for the various PNS models considered in this study.
}
\begin{tabular}{ccccc}
\hline\hline
  EOS & Model & $c_0$ (kHz) & $c_1$ (kHz) & $c_2$ (kHz)  \\
\hline
  Shen & W40 & 0.4525  & 1.4453 & $-0.1618$  \\
            & T50 &  0.3960  & 1.7183 & $-0.3943$  \\
 LS180 & W40 & 0.4629 & 0.9244 & 4.2095  \\
           &  T50 &  0.5018  & 0.1830 & 6.6968  \\
 LS220 & W40 & 0.3181 & 2.6230 & 0.3272  \\
\hline\hline
\end{tabular}
\label{tab:fit_fft}
\end{table}

\begin{figure}[tbp]
\begin{center}
\includegraphics[scale=0.5]{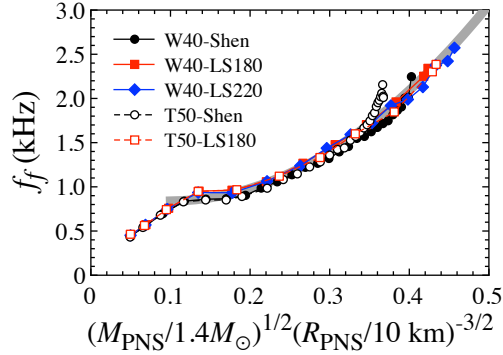} 
\end{center}
\caption{
Frequencies of the $f$-mode gravitational waves from various PNS models are shown as a function of the square root of the PNS average density for each time step. The solid-thick line denotes the fitting formula given by Eq. (\ref{eq:fit_ffave}), using the data for PNSs with the square root of the average density larger than $0.1$.
}
\label{fig:ffave}
\end{figure}

In addition, as in Fig~\ref{fig:ffave}, we find that the frequency of the $f$-mode gravitational waves strongly depends on the square root of the PNS average density, $(M_{\rm PNS}/1.4M_\odot)^{1/2}(R_{\rm PNS}/10\,{\rm km})^{-3/2}$, almost independently of the progenitor models. This behavior is understood, because the $f$-mode gravitational waves are excited as a result of acoustic oscillations, which can be characterized by the average density. In fact, it is shown that the $f$-mode frequency can be written as a linear function of the square root of the average density not only for cold neutron stars \cite{AK1996,AK1998}, but also for the PNS born after the core-collapse supernovae at least for the early phase after core bounce (up to $\sim 1$ second)\cite{ST2016,SKTK2017,SKTK2019}. Owing to this behavior of the $f$-mode frequency, it is suggested that the evolution of the PNS average density could be determined via the direct observation of the $f$-mode gravitational waves. On the other hand, for the black hole formation considered in this study, we find the $f$-mode frequency can be fitted as  
\begin{equation}
  f_f {\rm (kH)} = 0.9733 - 2.7171  X + 13.7809  X^2, \label{eq:fit_ffave}
\end{equation}
where $X$ is the square root of the PNS average density defined by $X\equiv (M_{\rm PNS}/1.4M_\odot)^{1/2}(R_{\rm PNS}/10\,{\rm km})^{-3/2}$, i.e., fitting is not a linear function of $X$. In this way, the evolution of the PNS average density would be determined via the direct observation of the $f$-mode gravitational waves. That is, using the observed frequency of $f$-mode gravitational wave, $f_f(T_{\rm pb})$, the square root of the PNS average density, $X(T_{\rm pb})$, is determined by
\begin{equation}
   X(T_{\rm pb}) = 9.8582\times 10^{-2}\times\left[1 + \left\{1- 7.2673\times \left(1-\frac{f_f(T_{\rm pb})}{0.9733}\right)\right\}^{1/2} \right], 
    \label{eq:ave}
\end{equation}
where $f_f$ is in the unit of kHz.  We stress that the evolution of $f$-mode frequency for the PNSs that would collapse to black hole is different from that for ordinary PNSs.  It is possible to explore higher average density for the black hole forming case and the $f$-mode frequency evolves beyond the linear relation, which was used in the ordinary PNSs as in Refs.~\cite{ST2016,SKTK2017,SKTK2019}. 
We remark that the behavior of the $f$-mode frequency without the relativistic Cowling approximation may not be so simple for non-accreting PNSs, where the $f$-mode frequency would evolve non-monotonically~\cite{FMP2003,Camelio17}.

Furthermore, as an advantage for considering the black hole formation, we can discuss the PNS model with the maximum mass allowed by the adopted EOS. As mentioned above, the moment when the PNS would collapse to a black hole can be determined via the neutrino observation.  The PNS models at this moment should correspond to that with the maximum mass allowed by the adopted EOS. We remark that the maximum mass of the PNS is different from that for the cold neutron star even with the same nuclear EOS, due to the effects of the entropy and lepton distributions including neutrinos. As shown in Table~\ref{tab:fitR} and Fig.~\ref{fig:fft-fit}, $T_{\rm BH}$ and the evolution of $f$-mode gravitational waves depend on the progenitor models of PNSs and the EOS.  Even so,  the observed $f$-mode gravitational wave can be generally fitted as Eq.~(\ref{eq:fit_fft}). Hence,
via Eq.~(\ref{eq:fit_fft}) with the help of the neutrino observation, one can estimate the frequency of the $f$-mode gravitational waves from the last moment of the PNS re-collapse to a black hole. 
With the resultant $f$-mode frequency (or the observed $f$-mode frequency if the $f$-mode gravitational wave would be directly observed), the average density of the PNS with the maximum mass can be estimated via Eq.~(\ref{eq:ave}).

Now, we check how the scheme suggested above works, using the specific PNS models considered in this study. That is, how well one can determine the square root of the average density for the PNS with the maximum mass via Eqs.~(\ref{eq:fit_fft}) and (\ref{eq:ave}) with the help of the neutrino observation. In practice, we determine the square root of the average density for PNS with the maximum mass estimated with this scheme, $X_{\rm est}$. Then, we compare it with the corresponding value, $X_{\rm sim}$, which is known from the simulation in advance. The result is shown in Table~\ref{tab:error} for various PNS models, where the relative error determined by $(X_{\rm est} - X_{\rm sim})/X_{\rm sim}$ is also shown. From this Table, we can find that the relative error in the square root of the average density for the PNS model with the maximum mass becomes at least less than $10\%$ independently of the progenitor models. In fact, one can estimate the square root of the average density for the PNS model with the maximum mass within less than $10\%$ accuracy even for the model of T50-Shen, for which the $f$-mode frequency deviates relatively from the fitting formula given by Eq.~(\ref{eq:fit_ffave}) as shown in Fig.~\ref{fig:ffave}.

\begin{table}
\centering
\caption{
Comparison between the square root of the average density of PNS with maximum mass estimated with the fitting formula (\ref{eq:fit_fft}) and the neutrino observation, $X_{\rm est}$, and the corresponding value, $X_{\rm sim}$, which is known from the simulation in advance. The relative error is calculated by $(X_{\rm est}-X_{\rm sim})/X_{\rm sim}$.
}
\begin{tabular}{ccccc}
\hline\hline
  EOS & Model & $X_{\rm est}$ & $X_{\rm sim}$ & relative error ($\%$)  \\
\hline
  Shen & W40 & 0.4014  & 0.4027 & $-0.32$  \\
            & T50 & 0.4000   & 0.3674 & $8.87$  \\
 LS180 & W40 & 0.4230  & 0.4234 & $-0.09$  \\
           &  T50 &  0.4244  & 0.4336 &  $-2.12$ \\
 LS220 & W40 & 0.4532 & 0.4567 &  $-0.77$ \\
\hline\hline
\end{tabular}
\label{tab:error}
\end{table}

\begin{figure*}[tbp]
\begin{center}
\begin{tabular}{cc}
\includegraphics[scale=0.5]{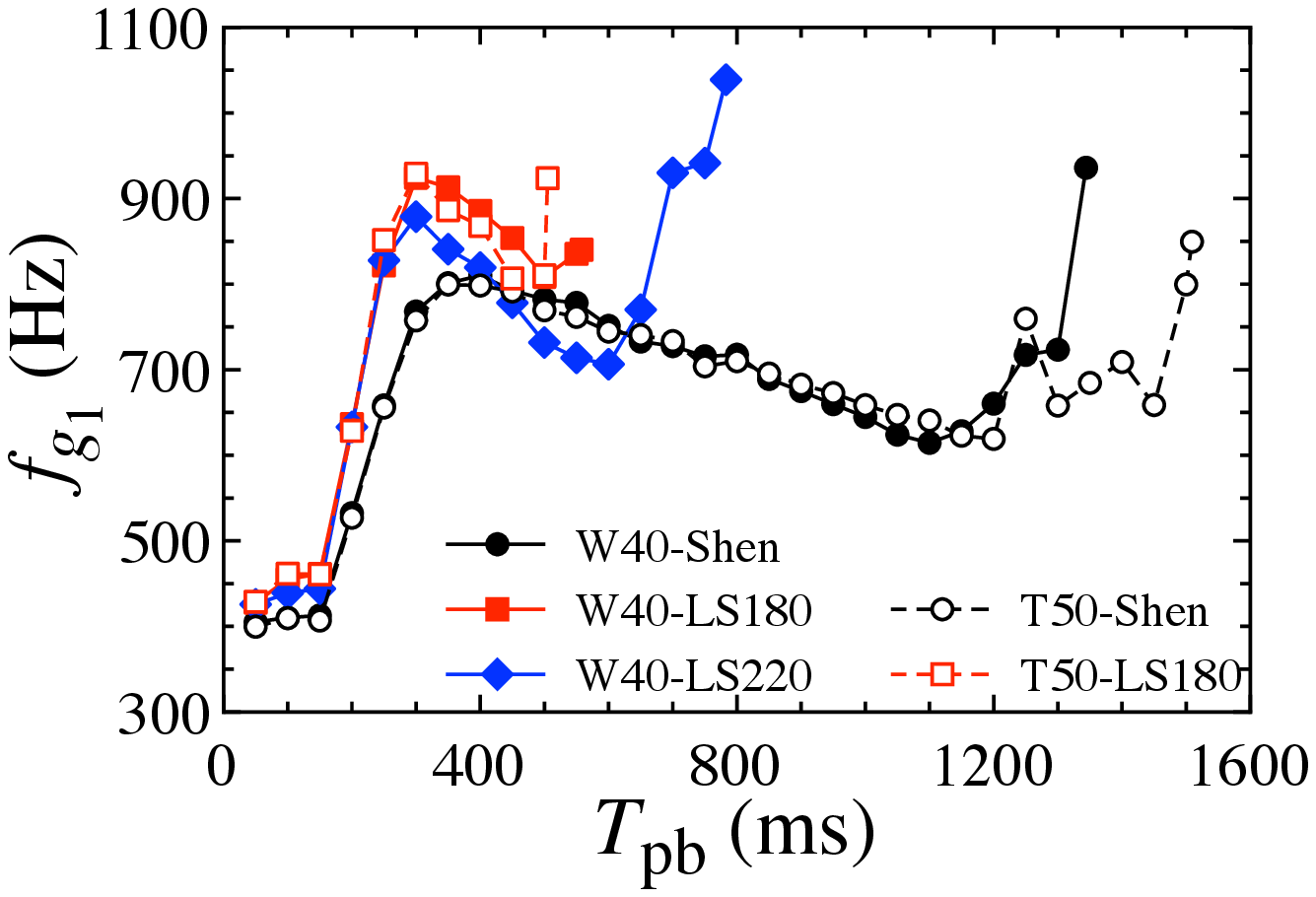} &
\includegraphics[scale=0.5]{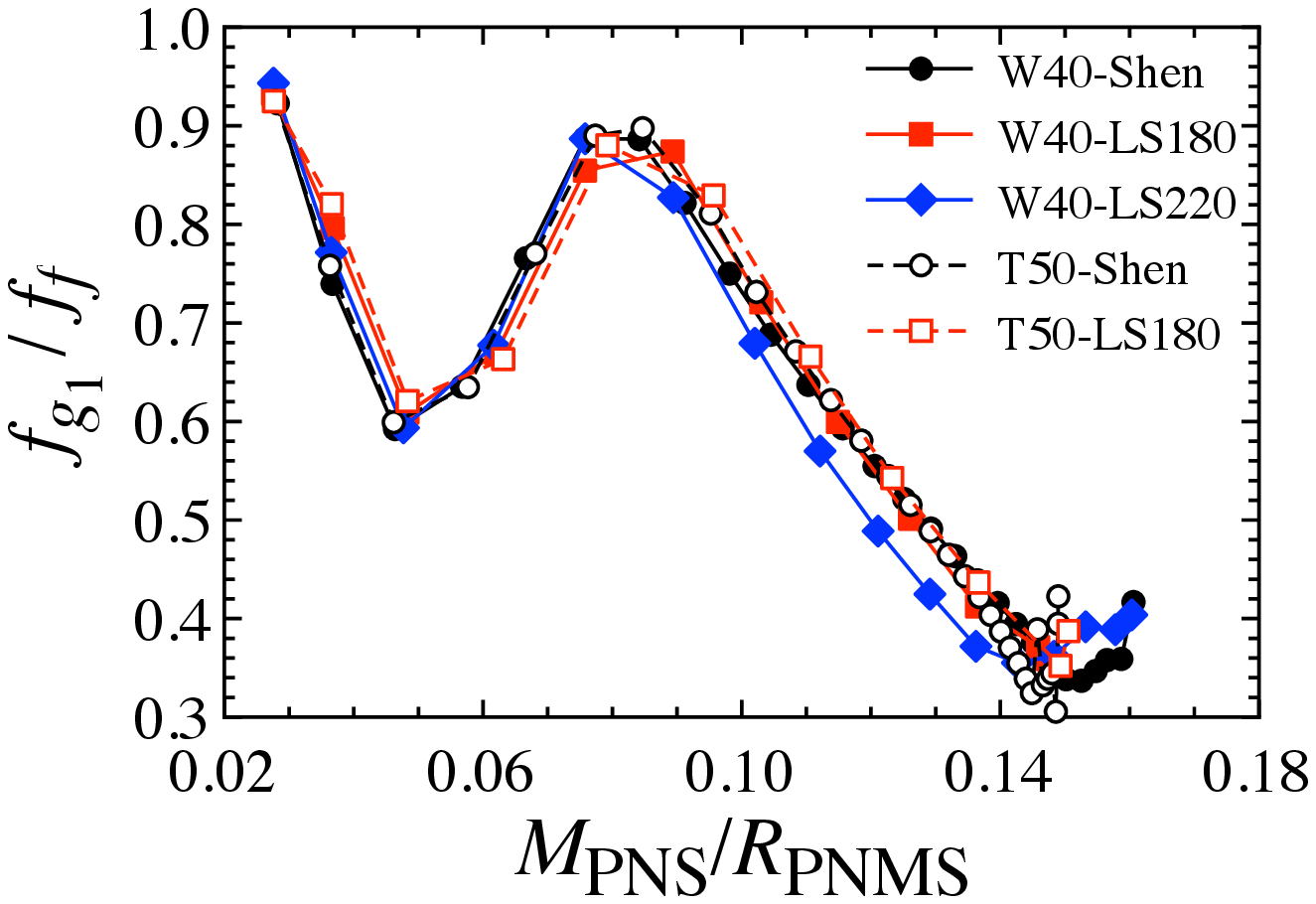}  
\end{tabular}
\end{center}
\caption{
Left: evolution of the frequency of the $g_1$-mode gravitational wave for various PNS models. Right: ratio of the $g_1$-mode frequency to the $f$-mode frequency, $f_{g_1}/f_f$, as a function of the PNS compactness.
}
\label{fig:fgff}
\end{figure*}

In contrast to the $f$-mode frequency, one can see the complex behavior in the $g_i$-mode frequency. It comes from the avoided crossing with the $f$- and $g_2$-modes for $i=1$ and with the $g_{i\pm1}$-modes for $i>1$. For example, as shown in Fig.~\ref{fig:f-W40Shen}, the avoided crossing in the $g_1$-mode frequency for the PNS model with W40-Shen happens at $T_{\rm pb}\sim (300-350)$ ms with $f$-mode and at $T_{\rm pb}\sim 150$ and $\sim 1100$ ms with $g_2$-mode. For various progenitor models, we show the evolution of the $g_1$-mode gravitational waves in the left panel of Fig.~\ref{fig:fgff}, from which one can observe that the time evolution of the $g_1$-mode gravitational waves strongly depends on the progenitor models. In particular, it seems that the time when the avoided crossing happens between the $g_1$- and $g_2$-mode gravitational waves strongly depends on the progenitor models.

Nevertheless, we find the universal behavior, which is insensitive to the progenitor models, in the ratio of the $g_1$-mode frequency to the $f$-mode frequency, $f_{g_1}/f_f$, as shown in the right panel of Fig.~\ref{fig:fgff}. That is, $f_{g_1}/f_f$ can be characterized well by the PNS compactness independently of the progenitor models, although the progenitor dependence of $f_{g_1}/f_f$ remains in the final phase just before the PNS would collapse to a black hole \cite{p/f}. Owing to this feature, one can derive the PNS compactness at each time step via the observation of the $f$- and $g_1$-mode gravitational waves. On the other hand, from the observation of the $f$-mode gravitational waves, one can derive also the PNS average density, as mentioned above. Thus, via the simultaneous observation of the $f$- and $g_1$-mode gravitational waves from the PNSs, one can determine the evolution of the PNS average density and compactness independently of the progenitor models. With this information, one can reconstruct the evolution of the PNS radius and mass, which enables us to probe the EOS for a high density region.

\section{Conclusion and discussion}
\label{sec:Conclusion}

We examine the eigenfrequency of the accreting PNSs toward the black hole formation for the two progenitor models with three EOSs.  As PNS models, we adopt the numerical results of spherically symmetric neutrino-radiation hydrodynamics in general relativity.  We focus on the black hole forming case having the information of time evolution of neutrino emission with the termination of neutrino burst at the black hole formation.  We performed a linear perturbation analysis by assuming the relativistic Cowling approximation. We show the evolution of the eigenfrequencies with the occurrence of the avoided crossing between the adjoining modes.  We find that the frequency of the $f$-mode gravitational waves can be expressed as a function of the PNS average density independently of the progenitor models. We also find that the ratio of the $g_1$-mode frequency to the $f$-mode frequency can be characterized by the PNS compactness independently of the progenitor models. As a result, it is found that the evolution of the PNS average density and compactness can be determined independently of the progenitor models via the simultaneous observations of the $f$- and $g_1$-mode gravitational waves. In addition, we can suggest that the average density of the PNS with the maximum mass could be determined via the observations of the $f$-mode gravitational waves together with the neutrino observation. That is, the time when the PNS would collapse to a black hole will be provided from the neutrino observation, while the PNS average density at such a time could be determined from the universal relation between the average density and the $f$-mode frequency. These information for the PNS could help us to constrain the EOS for a high density region.

Some of the numerical simulations (e.g., Ref.~\cite{KKT2016}) show not only the $f$-mode like gravitational wave signal (the so-called surface $g$-mode) but also the gravitational wave signal with low frequency ($\sim 100$ Hz). This low frequency signal is considered as a result from the standing accretion-shock instability (or maybe the $g$-mode oscillations of PNS), but the origin of it is still uncertain. If one would examine the ratio of low frequency signal to the $f$-mode like signal with various progenitor models and show such a ratio is almost insensitive to the progenitor models, one may be able to conclude that low frequency signal is strongly associated with the $g$-mode oscillations.

In this study, we performed the linear analysis, which does not provide information on the energy that excites each oscillating mode. Even so, the radiation energy was estimated via several numerical simulations (e.g., \cite{KKT2016,Kuroda17}). These results show the ramp-up mode as primary gravitational wave signals, which may correspond to the $f$-mode oscillations. The strain of this signal is estimated to be $h \sim 10^{-22}-10^{-23}$ for a Galactic event at a distance of $\sim 10$ kpc.  Assuming that the energy of ramp-up mode comes from the $f$-mode gravitational waves, one may detect this oscillating mode for a Galactic event even via the current gravitational wave detectors. In addition, the low frequency signal may be considered as a secondary signal, whose strain is estimated to be $h \sim 10^{-23}$. In this case, if this signal corresponds to the $g$-mode oscillations, the detection with the current detectors may be not possible even for a Galactic event. Feasibility of reconstruction of gravitational wave form for a nearby event has been explored in Ref.~\cite{Nakamura16}.  
Note that the magnitude of gravitational wave is still uncertain depending on the appearance of hydrodynamical instabilities in the shock dynamics found in numerical simulations (e.g.,~\cite{MJM2013,Nakamura16}, for an estimate of energy of gravitational wave from supernova explosions and observation strategy of multi-messenger signals.)

Last but not least, it would be beneficial to perform further systematic studies with progenitors and EOSs in order to firmly establish the analytic formula(e) and its application to observations although we believe that our finding holds in general.  
Note that the relation may not be simply explained by Eq.~(\ref{eq:fit_ffave}) if the $f$-mode gravitational waves without Cowling approximation evolve non-monotonically as shown in Refs.~\cite{FMP2003,Camelio17} for non-accreting PNSs.
It is also interesting to analyze the multi-messenger signals in multi-dimensional simulation of black hole forming cases (See Ref.~\cite{Pan18,Kuroda18} for example) since the current analysis is done under the spherical symmetry.  



\acknowledgments
This work is supported in part by
Grant-in-Aid for Scientific Research
(JP26104006, JP15K05093, JP17K05458, JP18H01212, JP19K03837)
and 
Grant-in-Aid for Scientific Research on Innovative areas 
"Gravitational wave physics and astronomy:Genesis"
(JP17H06357, JP17H06365)
from the Ministry of Education, Culture, Sports, Science and Technology (MEXT), Japan 

For providing high performance computing resources, 
Computing Research Center, KEK, 
JLDG on SINET4 of NII, 
Research Center for Nuclear Physics, Osaka University, 
Yukawa Institute of Theoretical Physics, Kyoto University, 
Nagoya University, Hokkaido University, 
and 
Information Technology Center, University of Tokyo are acknowledged. 

This work was partly supported by 
research programs at K-computer of the RIKEN AICS, 
HPCI Strategic Program of Japanese MEXT, 
"Priority Issue on Post-K computer" (Elucidation of the Fundamental Laws and Evolution of the Universe)
and 
Joint Institute for Computational Fundamental Sciences (JICFus).

\appendix
\section{Evolutions of the eigenfrequencies for various progenitor models}   
\label{sec:appendix_1}

\begin{figure*}[htbp]
\begin{center}
\begin{tabular}{cc}
\includegraphics[scale=0.5]{W40Shen-1} &
\includegraphics[scale=0.5]{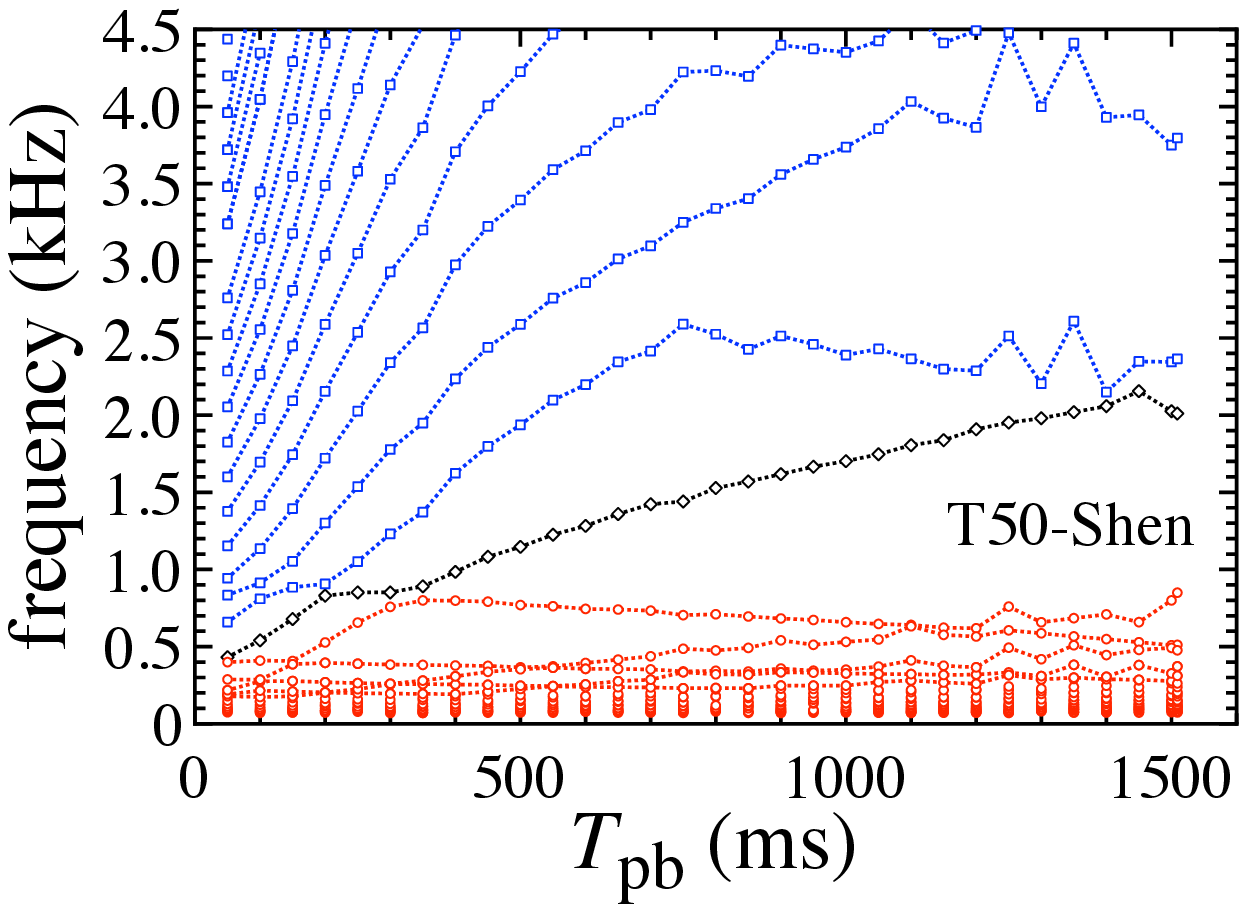} \\
\includegraphics[scale=0.5]{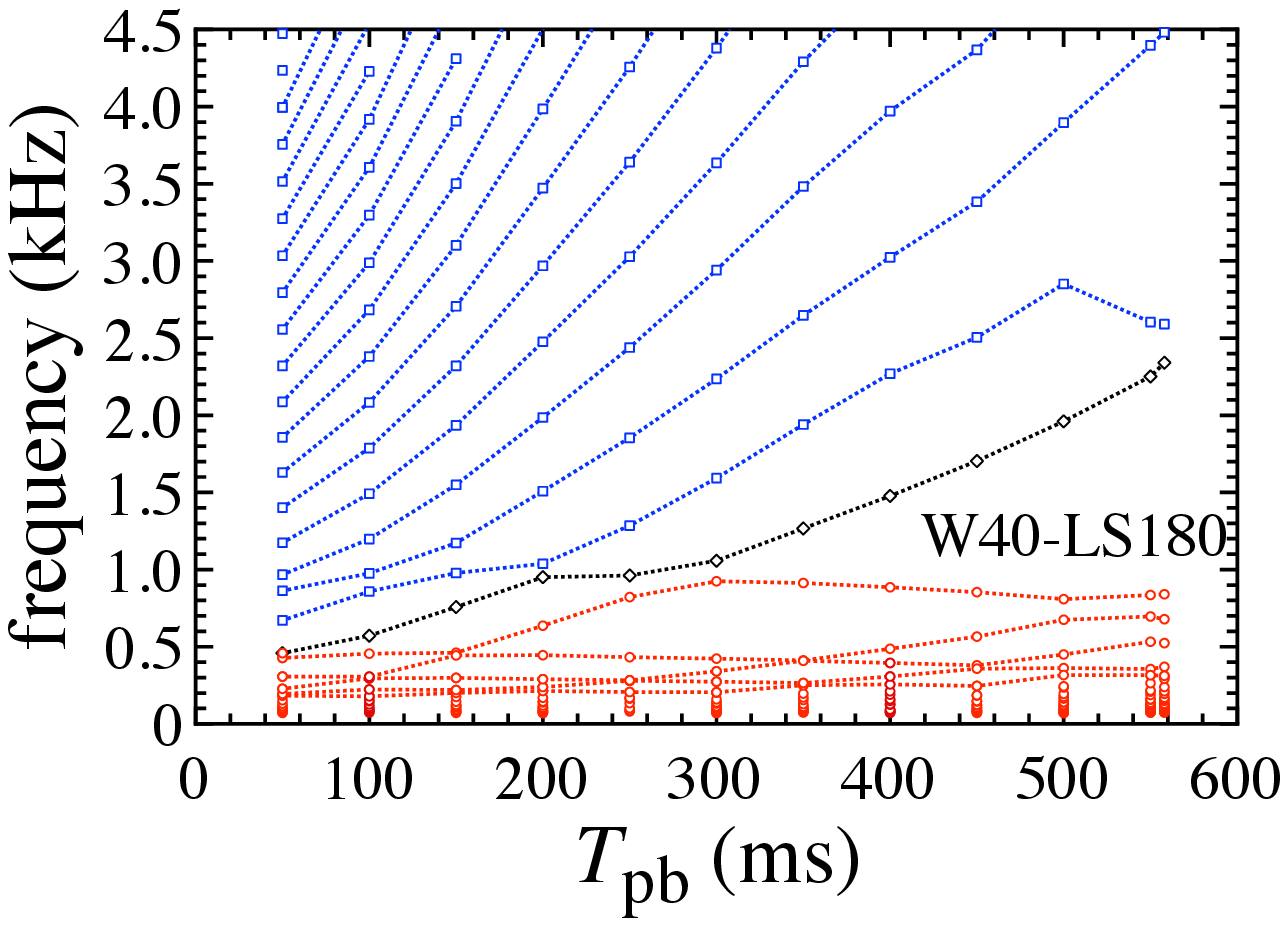} &
\includegraphics[scale=0.5]{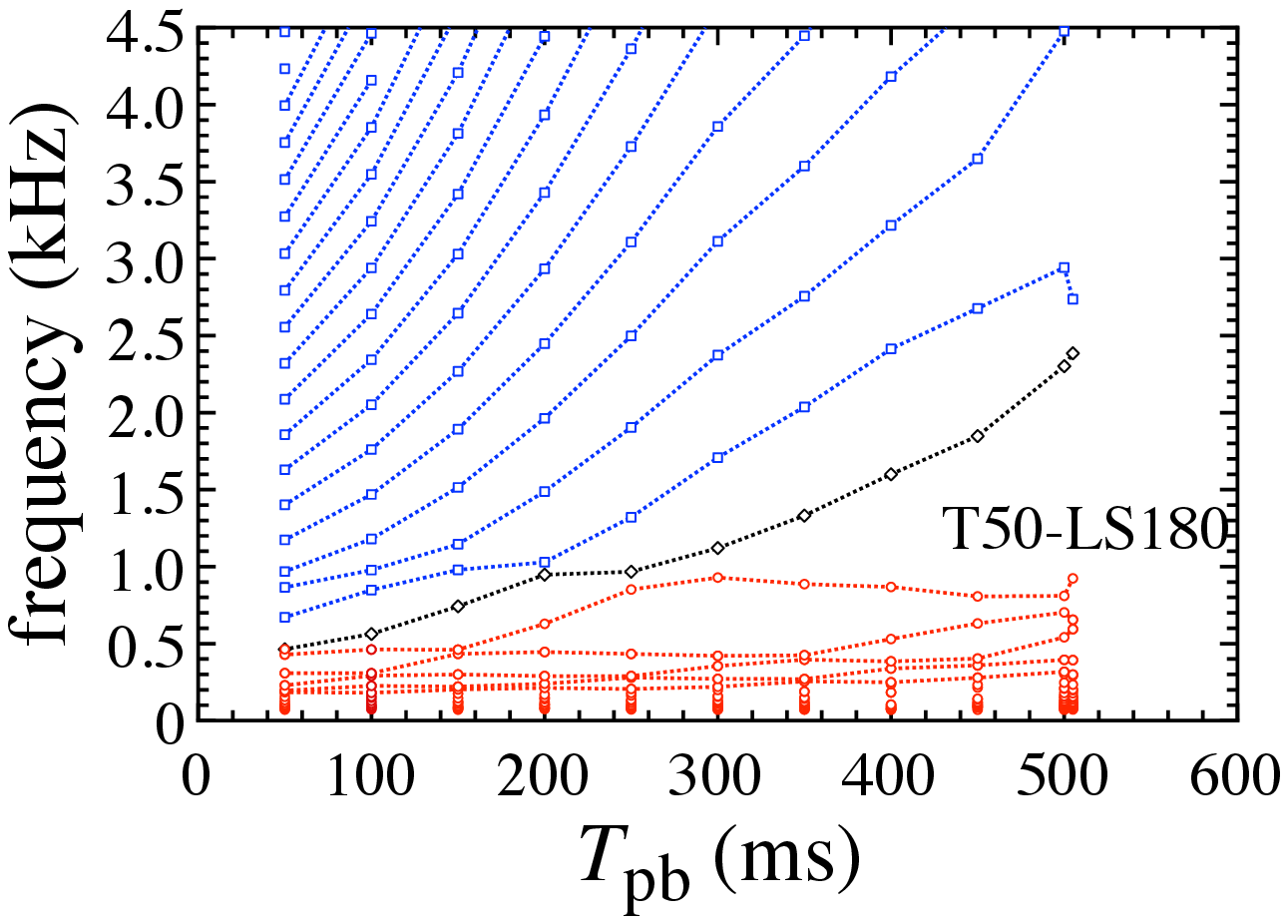} \\
\includegraphics[scale=0.5]{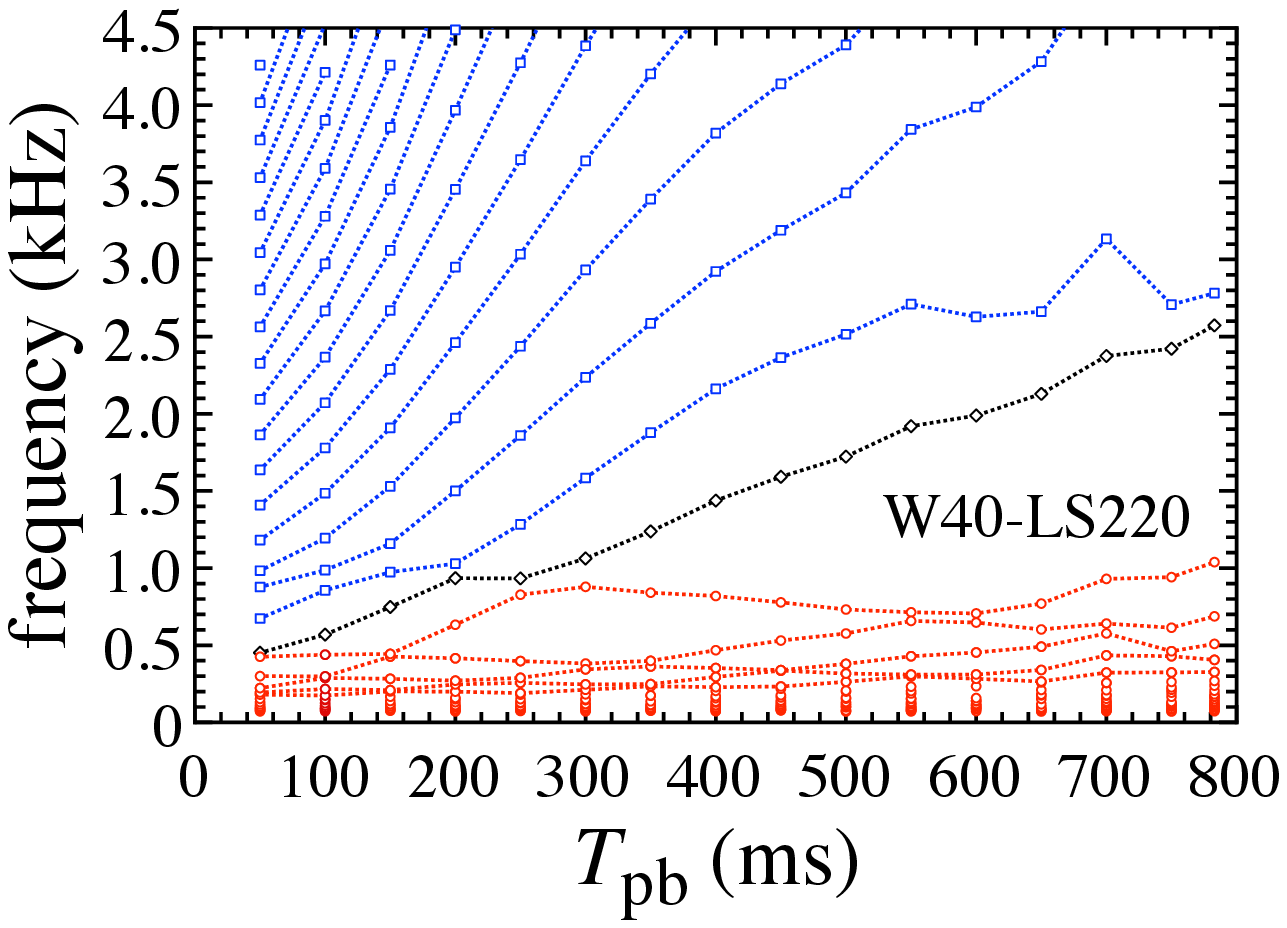} &  
\end{tabular}
\end{center}
\caption{
Evolutions of the eigenfrequencies from the PNS models produced by the various progenitor models. The left and right panels correspond to the results with W40 and T50, respectively. The top, middle, and bottom panels correspond the PNS constructed with Shen, LS180, and LS220, respectively. That is, the top-left panel is the same as the left panel of Fig.~\ref{fig:f-W40Shen}.
}
\label{fig:eigen}
\end{figure*}

In Fig.~\ref{fig:eigen}, we show the evolutions of the eigenfrequencies for various progenitor models. In all panels, the diamonds, squares and circles correspond to the $f$-, $p_i$-, and $g_i$-mode gravitational waves. The qualitative behavior is almost the same as each other, where one can observe the avoided crossing as in Fig.~\ref{fig:f-W40Shen}. As mentioned in text, one can also see that  the time when the avoided crossing happens between the $g_1$- and $g_2$-modes depends on the progenitor models. In addition, the behavior of some eigenfrequencies seems to be a little strange for the late phase, which may come from the less resolution in the numerical simulation with the baryon mass coordinate.

\section{Dependence of the frequency ratio of $f$-mode to $p_1$-mode}   
\label{sec:appendix_2}

\begin{figure}[tbp]
\begin{center}
\includegraphics[scale=0.5]{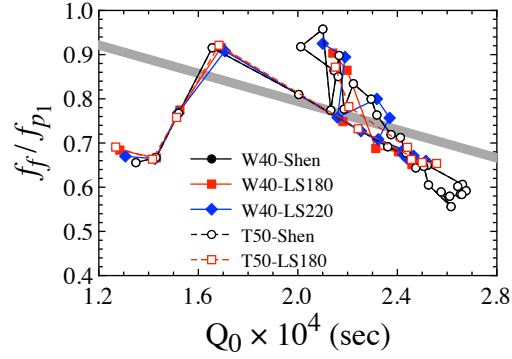} 
\end{center}
\caption{
The frequency ratio of the $f$-mode to the $p_1$-mode is shown as a function of $Q_0$ given by Eq.~(\ref{eq:Q0}) for the PNS models discussed in this study. The thick-solid line denotes the fitting line (Eq. (\ref{eq:ratio_Q0})) derived in Ref.~\cite{Camelio17}. 
}
\label{fig:ratio_Q0}
\end{figure}

In Ref.~\cite{Camelio17}, it has been shown that the frequency ratio of the $f$-mode to the $p_1$-mode is also strongly associated with the PNS average density, almost independently of the progenitor mass and EOS.  In fact, introducing a new property, $Q_0$, defined by 
\begin{equation}
  Q_0 = \frac{1}{f_f}\left(\frac{M_{\rm PNS}}{M_\odot}\right)^{1/2}
     \left(\frac{R_{\rm PNS}}{10\ {\rm km}}\right)^{-3/2}, \label{eq:Q0}
\end{equation}
the fitting formula was derived~\cite{Camelio17}, such as
\begin{equation}
  \frac{f_f}{f_{p_1}} = 1.1131 - 1596\times\left(\frac{Q_0}{1\ {\rm sec}}\right). \label{eq:ratio_Q0}
\end{equation}
Since it would be interesting to check whether or not this relation is still held with our results discussed in this study, we plot $f_f/f_{p_1}$ for various PNS models as a function of $Q_0$ in Fig.~\ref{fig:ratio_Q0}, where we also show the relation given by Eq.~(\ref{eq:ratio_Q0}) with the thick-solid line for reference. From this figure, we find that $f_f/f_{p_1}$ calculated with our results is more or less associated with $Q_0$ independently of the PNS models, although the correlation is weaker than that shown in Ref.~\cite{Camelio17}.

\begin{thebibliography}{999}
\bibitem{p/f}
   It has been reported that the ratio of the $p_1$-mode to the $f$-mode gravitational waves from the PNSs strongly depends on the square root of the PNS average density, which is almost independent of the EOS and progenitor mass \cite{Camelio17}. With our results for the various PNSs discussed in this study, we check whether or not such a correlation is still held in Appendix \ref{sec:appendix_2}.
\end{thebibliography}
\end{document}